\documentclass[aps,prd,reprint,showkeys,nofootinbib,superscriptaddress,notitlepage]{revtex4-1}
\usepackage{amsmath,graphicx,amssymb,multirow,url,xspace}
\usepackage[usenames,dvipsnames]{xcolor} 
\usepackage{aas_macros}
\usepackage[utf8]{inputenc}
\usepackage[unicode=true,
  bookmarks=false,   backref=false,    colorlinks=true,
  linktocpage=true,  citecolor=black,  linkcolor=black,
  urlcolor=black,    breaklinks=true
]{hyperref}

\newcommand{\GN}{G_{\rm N}}
\newcommand{\relG}{\Delta G/\GN}
\newcommand{\relGmax}{\relG |_\mathrm{max}}
\newcommand{\Rmax}{R_\mathrm{max}}
\newcommand{\mpc}{\mathrm{Mpc}}
\newcommand{\hu}{\mathrm{km/s/Mpc}} 
\newcommand{\MHW}{M_{H}^W}
\newcommand{\sy}{\scriptstyle}

\newcommand{\td}{..}
\newcommand{\oh}{{\rm [O/H]}}
\newcommand{\p}{{[P]}}

\newcommand{\D}{\hyperlink{Desmond_2019}{D19}\xspace}
\newcommand{\Fold}{\hyperlink{Freedman_2019}{F19}\xspace}
\newcommand{\Fnew}{\hyperlink{Freedman:2024eph}{F25}\xspace}
\newcommand{\A}{\hyperlink{Anand:2021sum}{A22}\xspace}
\newcommand{\Lnew}{\hyperlink{Li:2024pjo}{L24}\xspace}

\newcommand{\be}{\begin{equation}}
\newcommand{\ee}{\end{equation}}

\begin{document}

\title{
Constraining Fifth Forces using the Local Distance Ladder: Implications for the Hubble Tension
}

\author{Marcus \surname{Högås}}
\email{marcus.hogas@fysik.su.se}
\thanks{corresponding author}
\affiliation{Oskar Klein Centre, Department of Physics, Stockholm University, Albanova University Center, 106 91 Stockholm, Sweden}

\author{Edvard \surname{Mörtsell}}
\email{edvard@fysik.su.se}
\affiliation{Oskar Klein Centre, Department of Physics, Stockholm University, Albanova University Center, 106 91 Stockholm, Sweden}

\author{Harry \surname{Desmond}}
\email{harry.desmond@port.ac.uk}
\affiliation{Institute of Cosmology \& Gravitation, University of Portsmouth, Dennis Sciama Building, Portsmouth, PO1 3FX, UK}

\author{Adam \surname{Riess}}
\email{ariess@stsci.edu}
\affiliation{Space Telescope Science Institute, Baltimore, MD 21218, USA}
\affiliation{Department of Physics and Astronomy, Johns Hopkins University, Baltimore, MD 21218, USA}
 
\begin{abstract}
We revisit the local distance ladder measurement of the Hubble constant in models where gravity is modified by a fifth force, an additional long-range interaction. In many such theories the force is screened—suppressed in dense environments but potentially active in galaxies used for distance calibration. We model this environmental dependence using three quantities that characterize each galaxy’s large-scale gravitational environment: the external gravitational potential $\Phi$, acceleration $a$, and curvature $K$. 
Our baseline analysis recalibrates the SH0ES-team's Cepheid–supernova distance ladder, incorporating the fifth force via its impact on the Cepheid period–luminosity relation. Across models, a fifth force is strongly constrained, with posteriors concentrated around a null result. The inferred Hubble constant is $H_0 = 73.1 \pm 1.0 \, \mathrm{km/s/Mpc}$, retaining the Hubble tension at $>5 \, \sigma$.
As an additional test, we incorporate four independent Tip of the Red Giant Branch (TRGB) distance datasets into a joint Cepheid–TRGB–supernova calibration. These combined analyses further constrain the magnitude of  fifth-force effects.
Taken together, our results show that, across the class of screened fifth-force models we analyze, the calibration of the local distance ladder remains essentially unchanged, leaving the Hubble tension intact.
\end{abstract}

\maketitle

\section{Introduction}
High-precision measurements of the Hubble constant, $H_0$, have revealed a persistent discrepancy between early- and late-Universe determinations. Observations of the cosmic microwave background (CMB) by the \emph{Planck} satellite yield $H_0 = 67.4 \pm 0.5 \, \hu$ when interpreted within the $\Lambda$CDM model \cite{Planck2020}. In contrast, the local distance ladder calibrated with Cepheids and Type Ia supernovae (SNe~Ia) yields a significantly higher value, $H_0 = 73.0 \pm 1.0 \, \hu$, as measured by the SH0ES\footnote{Supernovae and $H_0$ for the Equation of State of dark energy.} collaboration \cite{Riess:2021jrx,H0DN:2025lyy}. This $5 \, \sigma$ discrepancy, known as the Hubble tension, remains one of the most prominent challenges in contemporary cosmology.

One proposed resolution is that gravity may deviate from general relativity (GR) on galactic scales. Many extensions of GR, with scalar–tensor theories providing the canonical example, predict an additional long-range interaction—often termed a fifth force—whose strength depends on the surrounding gravitational environment. Such forces must be screened in high-density regions to satisfy solar-system tests, but may remain active in lower-density environments where standard-candle distance indicators reside. If active, they can systematically shift the intrinsic luminosities of these distance indicators, altering the calibration of the local distance ladder and thereby the inferred value of $H_0$ \cite{Desmond_2019,Sakstein:2019qgn,Desmond:2020wep,Hogas:2023vim,Hogas:2023pjz}.

In this paper we revisit this possibility and extend previous studies in several ways:
\begin{itemize}
    \item We perform a full Bayesian recalibration of the distance ladder using the fourth-iteration SH0ES dataset as our baseline \cite{Riess:2021jrx}. This more than doubles the number of SN~Ia host galaxies relative to earlier iterations \cite{Riess:2009pu,Riess:2011yx,Riess:2016jrr} and explicitly incorporates the full covariance structure of both the Cepheid and SN~Ia data.

    \item We optionally include the TRGB distances together with the Cepheids and SNe~Ia in a single, uniform calibration, rather than treating them as an external cross-check.

    \item We propagate the uncertainty in the Cepheid instability-strip crossing (second versus third crossing) through the entire distance-ladder calibration.
\end{itemize}

\noindent \textbf{Notation.} Following Ref.~\cite{Desmond_2019} (\D hereafter), the externally sourced gravitational potential ($\Phi$) is given in units of $c^2$ where $c$ is the speed of light, the externally sourced acceleration ($a$) is given in units of $\mathrm{km/s^2}$, and the externally sourced curvature ($K$) is given in units of $1/\mathrm{cm}^2$. Throughout this work, ``$\log$'' denotes the base-10 logarithm. Concerning galaxy names, N4258 stands for NGC 4258 and U9391 stands for UGC 9391, etc.

\section{Theory}
In the weak-field, quasi-static limit---describing for example the gravitational field around a star or a galaxy---many extensions of GR predict departures from Newtonian gravity, effectively strengthening the gravitational interaction.
Scalar–tensor theories, in which a scalar mediates the additional interaction, are the canonical examples.
Although the fifth force generically takes a Yukawa form, it becomes an additional inverse-square contribution to Newtonian gravity when the scalar’s Compton wavelength exceeds the size of the system. In this regime it is convenient to describe the modification as a rescaling of the gravitational constant, introducing an effective gravitational constant,
\begin{equation}
    G = \GN + \Delta G
\end{equation}
where $\Delta G$ quantifies the strength of the fifth force and may vary with the environment.

Any viable fifth force must evade the stringent bounds from laboratory and solar-system tests, which constrain deviations from GR to extremely small levels in those environments \cite{Will:2014kxa}. At the same time, for such modifications to be relevant to astrophysics, they must produce observable deviations on larger scales or in different gravitational environments. This tension is resolved in many theories through screening mechanisms: nonlinear effects that suppress the scalar field in dense regions while allowing it to influence dynamics in weak-gravity environments. Examples include the chameleon, symmetron, and Vainshtein mechanisms \cite{Khoury:2013tda,Joyce:2014kja,Brax:2021wcv}.
A key consequence of screening is that the fifth force need not be active in all galaxies or in all regions within a galaxy. Instead, its presence depends sensitively on the surrounding gravitational environment. For further information on screened modified gravity see \cite{Clifton,NPP,Brax:2021wcv} and references therein.

To capture this environmental dependence in a computationally efficient way, we model the fifth-force strength using three alternative proxy fields: the gravitational potential $\Phi$, the acceleration $a$, and the curvature $K$. Each proxy corresponds to a different class of screened fifth-force theories and has been shown to trace fifth-force behavior accurately across a wide range of screening mechanisms \cite{Khoury:2013tda,Joyce:2014kja}.
This proxy-field approach allows us to test screened fifth-force phenomenology without committing to any specific underlying theory.
The proxy fields are computed using the maps of Ref.~\cite{Desmond_2017}, which are constructed by combining galaxy-survey data with halo catalogs from $N$-body simulations and numerical models of structure formation \cite{Lavaux:2015tsa,Jasche:2018oym}.
The proxy-field value for a specific galaxy receives contributions from all sources within a certain cutoff radius ($\Rmax$) which designates the range of the fifth force.

In regions where the proxy field falls below the screening threshold—treated as a free parameter in our model—the scalar field becomes unscreened and mediates a nonzero fifth force. Conversely, large values of the proxy field corresponds to screened regions in which $G = \GN$.
It is important to bear in mind that $\Phi$, $a$ and $K$ capture the screening behaviour of only a subset of theories. In particular they describe \emph{environmental} screening, caused by mass surrounding a galaxy rather than within it. Separate investigation would be required for the other screening models considered in e.g. \cite{Desmond_2019,Sakstein:2019qgn,Desmond:2020wep}, as we discuss further in Section~\ref{sec:Discussion}.

\section{Data}
All data required to reproduce this work are available at \cite{DistLadFifthForceGitHub}, the GitHub repository associated with this paper.
This provides both the data products generated in this study as well as external datasets compiled from other sources.

\begin{figure*}[t]
    \centering
	\includegraphics[width=\linewidth]{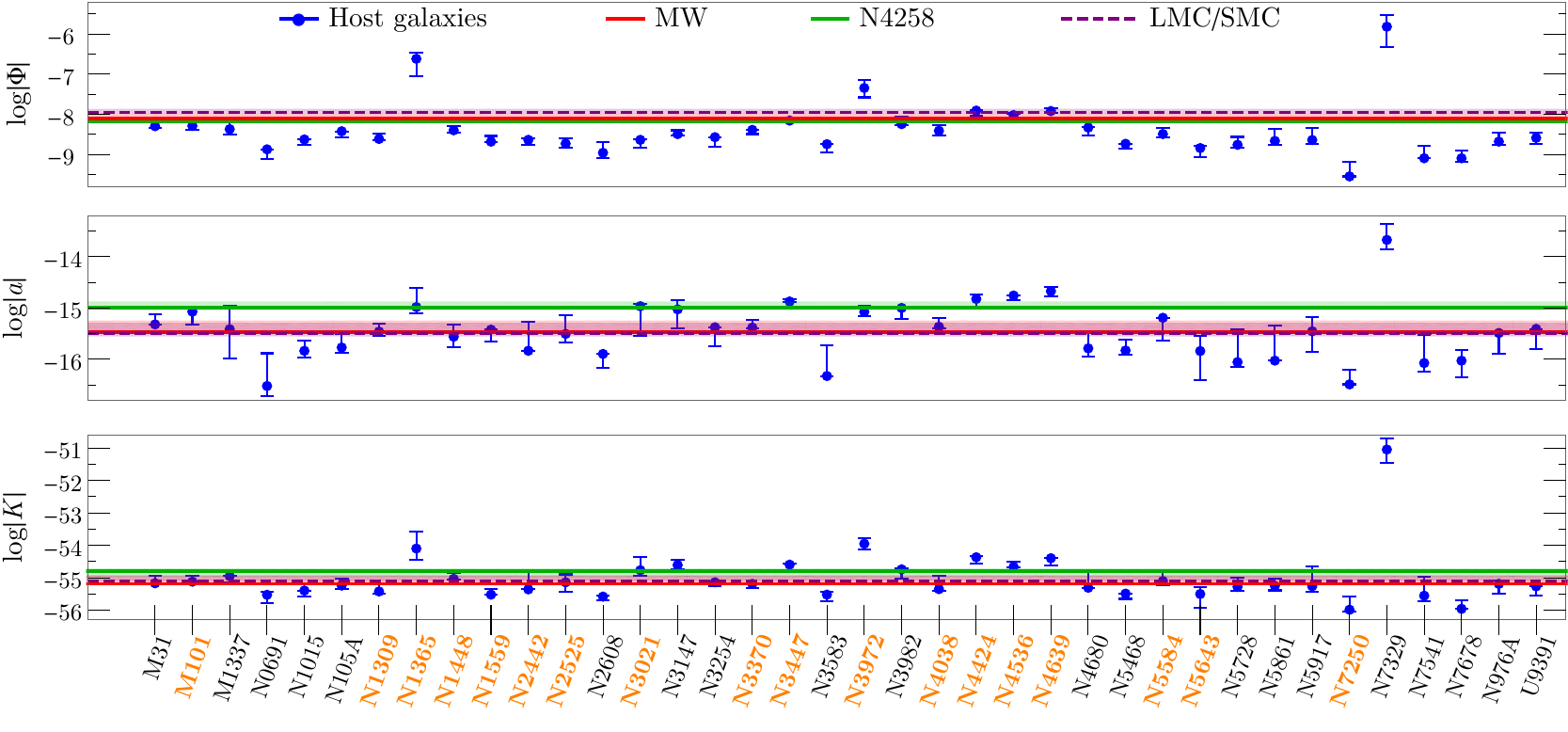}
    \caption{Proxy-field values for the galaxies used in the SH0ES distance ladder: the anchor galaxies, whose geometric distances calibrate the Cepheid period--luminosity relation, and SN~Ia host galaxies. Some of the galaxies also have TRGB-based distance estimates, and are marked in bold orange. The MW is shown in red, N4258 in green, and LMC/SMC in purple\footnote{LMC and SMC, being only $\simeq 0.02 \, \mpc$ apart, share identical proxy values due to their inhabiting the same large-scale environment.} whereas the SN~Ia host galaxies appear as blue points. The three panels illustrate the three alternative proxy-field descriptions of the fifth-force strength for a common cutoff radius $\Rmax = 0.4 \, \mpc$. \emph{Top:} gravitational potential $\Phi$. \emph{Middle:} acceleration $a$. \emph{Bottom:} curvature $K$.
    \label{fig:screeningproxyvalues}}
\end{figure*}

The fifth-force proxy fields ($\Phi$, $a$, and $K$) are determined using the reconstruction of Ref.~\cite{Desmond_2017}, which models the large-scale matter distribution out to $200 \, \mpc$. This method yields a posterior distribution of the proxy fields at the position of each galaxy, reflecting the uncertainties in the underlying density field. 
In the main calibration, we adopt the median of this posterior for each proxy field. The influence of the proxy-field uncertainties—and the extent to which they may alter $H_0$—is quantified and discussed in Section~\ref{sec:results}.
The proxy-field value for a specific galaxy receives contributions from all sources within a certain cutoff radius ($\Rmax$) which designates the range of the fifth force. Some examples are shown in Fig.~\ref{fig:screeningproxyvalues}.
We compute these values for five different cutoff radii, $\Rmax= [0.4, 1.4, 5.1, 18.1, 50] \, \mpc$, tabulated in Tab.~\ref{tab:ProxyVal}.

\begin{table*}
  \centering
  \begin{tabular}{c|ccc|ccc|ccc|c}
    \hline \hline
    Field & $\Phi_\mathrm{low}$ & $\Phi_\mathrm{median}$ & $\Phi_\mathrm{high}$ & $a_\mathrm{low}$ & $a_\mathrm{median}$ & $a_\mathrm{high}$ & $K_\mathrm{low}$ & $K_\mathrm{median}$ & $K_\mathrm{high}$ & $\Rmax$ \\
    & & $(c^2)$ & & & $(\mathrm{km/s^2})$ & & & $(\mathrm{cm}^{-2})$ & & $(\mpc)$ \\
    \hline
    MW & $-8.13$ & $-8.10$ & $-7.98$ & $-15.47$ & $-15.47$ & $-15.24$ & $-55.18$ & $-55.18$ & $-54.92$ & $0.4$ \\
    MW & $-7.18$ & $-7.11$ & $-7.03$ & $-15.50$ & $-15.44$ & $-15.07$ & $-55.13$ & $-55.07$ & $-54.89$ & $1.4$ \\
    \hline
  \end{tabular}
  \caption{Proxy-field values. Columns $\Phi_\mathrm{low}$, $\Phi_\mathrm{median}$, $\Phi_\mathrm{high}$ list the base-10 logarithm of $\Phi$ (in units of $c^2$). The ``low'' and ``high'' values give the lower and upper bounds of the $68 \, \%$ credible interval from~\cite{Desmond_2017}. The corresponding columns for $a$ and $K$ are defined analogously. This table is available in its entirety in machine-readable form at \cite{DistLadFifthForceGitHub}.}
  \label{tab:ProxyVal}
\end{table*}

We base our analysis on the fourth data release of the SH0ES project \cite{Riess:2021jrx}, with the data publicly available at \cite{PantheonPlusData}. For convenience, we use the reformatted version of the same data provided in Refs.~\cite{Hogas:2024qlt,MarcusSH0ESGitHub}, organized to facilitate direct application in alternative calibrations.
The Cepheid-based distance ladder comprises three rungs, as described below. Note that despite presenting the rungs separately below, we perform a joint calibration of all three rungs, as described in detail in Section~\ref{sec:StatisticalMethods} and Appendix~\ref{sec:AppData}.\\

\noindent \textbf{First rung.} In the first rung, the Cepheid period–luminosity (PL) relation—which links the pulsation period of a Cepheid to its intrinsic luminosity—is calibrated by anchoring Cepheid distances to independent geometric measurements in so-called anchor galaxies. In the fourth iteration of the SH0ES calibration, there are three primary anchor galaxies: the Milky Way (MW), where parallax provide direct distance measurements \cite{Riess_2021}; the Large Magellanic Cloud (LMC), whose Cepheid distances are determined via detached eclipsing binary stars \cite{Pietrzy_ski_2019}; and N4258, a galaxy with a maser-determined distance \cite{Reid_2019}.

Cepheid variable stars follow a well-established PL relation describing how their intrinsic luminosity is correlated with their pulsation period \cite{Leavitt:1907,Leavitt:1912zz}. The PL relation is typically expressed
\begin{equation}
    \label{eq:PLR}
    m_{H,i}^W = \mu_i + \MHW + b_W \p_i + Z_W \oh_i
\end{equation}
where $m_{H,i}^W$ is the Wesenheit magnitude of Cepheid number $i$ \citep{Madore:1991yf,Riess:2021jrx}, $\mu_i$ its distance modulus, and $\oh_i$ its metallicity, quantified as $\oh_i = \log \left[ (\mathrm{O/H})_i / (\mathrm{O/H})_\odot \right]$.
$\p_i$ quantifies the Cepheid period as $\p_i = \log (P_i / 10 \, \mathrm{days})$.

Because the anchor distances $\mu_i$ are known from geometric measurements, inserting them into Eq.~\eqref{eq:PLR} allows the intercept $\MHW$ and the slopes $b_W$ and $Z_W$ to be fitted directly from the observed magnitudes, colors, periods, and metallicities of the anchor Cepheids.\\

\noindent \textbf{Second rung.} In the second rung, the calibrated PL relation is used to determine the distances, $\mu_j$, to galaxies hosting SNe~Ia and can thereby be used to calibrate the peak absolute magnitude of SNe~Ia. The SH0ES program includes 42 SNe~Ia occurring in 37 galaxies that also host Cepheids.
SNe Ia serve as standardizable candles, meaning their peak brightness is nearly uniform after correcting for color, the shape of its light curve, and its host-galaxy properties \cite{Scolnic:2021amr}. We denote by $m_{B,j}$ the standardized rest-frame B-band apparent magnitude of supernova $j$, that is, the observed peak magnitude after applying these corrections. This quantity relates to the distance modulus through
\begin{equation}
    m_{B,j} = \mu_j + M_B
\end{equation}
where $M_B$ is the standardized absolute magnitude.

The standard SH0ES distance ladder uses Cepheids to calibrate the SN~Ia absolute magnitude. As an optional extension, we also incorporate TRGB distances in a unified calibration with the Cepheid and SN Ia data. This combined analysis is a strong discriminator of screened fifth-force models, since TRGB- and Cepheid-based distances respond oppositely to a fifth force \cite{Sakstein:2019qgn}.
In the extended analysis, we consider four TRGB datasets, each used separately in its own unified Cepheid–TRGB–SN~Ia calibration. Two datasets come from the Carnegie–Chicago Hubble Program (CCHP): the original HST-based measurements of Ref.~\cite{Freedman_2019} (\Fold) and the JWST-based update of Ref.~\cite{Freedman:2024eph} (\Fnew). The remaining two datasets are the HST-based measurements of Ref.~\cite{Anand:2021sum} (\A) and their JWST-based extension from Ref.~\cite{Li:2024pjo} (\Lnew). The CCHP dataset of Ref.~\cite{Freedman_2019} uses the LMC as the anchor for calibrating the absolute TRGB magnitude, whereas the other datasets adopt N4258 as the anchor.\\

\noindent \textbf{Third rung.} The third rung consists of 277 SNe~Ia in the Hubble flow (redshift $z > 0.023$) which are used to determine the Hubble constant. 
A second-order Taylor expansion of the distance--redshift relation around $z=0$ gives, for the $k$th Hubble flow supernova,
\begin{equation}
    \label{eq:mBHFSN}
    m_{B,k} - 5 \log \left[ c z_k \left\lbrace 1 + \frac{1}{2} (1-q_0) z_k \right\rbrace \right] - 25 = M_B - 5 \log H_0,
\end{equation}
where $q_0$ is the deceleration parameter.

With $M_B$ calibrated in the second rung, the Hubble constant $H_0$ can be inferred from the observed magnitudes and redshifts of the Hubble-flow supernovae.
For the left-hand side of Eq.~\eqref{eq:mBHFSN}, we use the values tabulated by the SH0ES team \cite{Riess:2021jrx}. Since these were computed assuming a deceleration parameter of $q_0 = -0.55$, our analysis inherits this choice. However, the inferred value of $H_0$ is largely insensitive to $q_0$, so this assumption has negligible impact on our results \cite{Riess:2021jrx}.

\section{Influence of a fifth force}
\label{sec:Influence}
The calibration of the distance ladder follows the structure of the SH0ES analysis \cite{Riess:2021jrx}, with the key difference that we incorporate the effect of a fifth force. In an optional extension to our baseline analysis, we also include TRGB distances directly in the data vector so that TRGB, Cepheid, and SN~Ia distances are calibrated simultaneously. A fifth force affects the rungs in the following way.\\

\noindent \textbf{First rung.} The geometric distances to the anchor galaxies are unaffected by a fifth force. This does not rely on any assumption about whether the anchors are screened or unscreened; it follows simply from the fact that the underlying distance measurements are purely geometric and therefore independent of $G$. As discussed in Ref.~\cite{Hogas:2023pjz}, \emph{Gaia} parallaxes (MW), megamaser orbital dynamics (N4258), and detached eclipsing binaries (LMC) determine distances using geometric or kinematic observables in which any change in 
$G$ either has no effect on the geometric observable (MW and LMC) or is absorbed by a nuisance parameter (the central black-hole mass in N4258). The inferred anchor distances are therefore unchanged under a fifth force, regardless of the local screening level.

Although the anchor distances themselves are unchanged, the fifth force can still modify the luminosity of the Cepheids, thereby shifting the zero-point of the PL relation \cite{Desmond_2019}.
The modified PL relation can therefore be written
\begin{equation}
    \label{eq:PLR_shift}
    m_{H,i}^W = \mu_i + \MHW + \Delta M_{H,i}^W + b_W \p_i + Z_W \oh_i
\end{equation}
where
\begin{equation}
    \label{eq:deltaMHW}
    \Delta M_{H,i}^W = - 2.5 \left( \frac{A}{2} + B_i \right) \log \left( 1 + \frac{\Delta G}{\GN} \right)
\end{equation}
encodes the change in intrinsic luminosity induced by the fifth force.
A positive $\Delta G$ therefore corresponds to brighter Cepheids.

In Eq.~\eqref{eq:deltaMHW}, the coefficient $A$ quantifies the fifth-force effect on the stellar core and is fixed to $A = 1.3$ following \D. The coefficient $B_i$ describes the contribution from the stellar envelope and depends on the Cepheid’s mass and on whether it is on its second or third crossing of the instability strip.\footnote{Cepheids cross the instability strip multiple times, and their envelope properties differ on the second and third crossings. The first instability-strip crossing is very brief compared to the second and third and constitutes a negligible fraction of the observed population.} The values are listed in Tab.~\ref{tab:CephB}. The envelope contribution vanishes ($B_i = 0$) if the envelope is screened. In this work, to explore the maximal impact of a fifth force, we assume that the core and envelope are either both screened or both unscreened.
Because the fifth-force strength is modeled through proxy fields that trace the large-scale structure, we assume that $\relG$ takes a single, galaxy-wide value for all Cepheids within a given host. However, because $B_i$ varies from star to star, the resulting magnitude shifts $\Delta M_{H,i}^W$ still differ among Cepheids within the same galaxy.\\

\begin{table*}
  \centering
  \begin{tabular}{c|ccccccccc}
    \hline \hline
    & $5 M_\odot$ & $6 M_\odot$ & $7 M_\odot$ & $8 M_\odot$ &$9 M_\odot$ &$10 M_\odot$ &$11 M_\odot$ &$12 M_\odot$ &$13 M_\odot$\\
    \hline
    2nd crossing & 4.21 & 4.52 & 4.45 & 4.34 & 4.18 & 4.00 & 3.81 & 3.67 & 3.58 \\
    3rd crossing & 3.67 & 3.61 & 3.79 & 3.58 & 3.46 & 3.48 & 3.58 & 3.92 & 3.95 \\
    \hline \hline
  \end{tabular}
  \caption{Values for $B$ (Eq.~\ref{eq:deltaMHW}) depending on the mass of the Cepheid and whether it is on the second or third crossing of the instability strip. Reproduced from Ref.~\cite{Sakstein:2019qgn}.}
  \label{tab:CephB}
\end{table*}

\noindent \textbf{Second rung.} As we have seen, a fifth force induces a shift in Cepheid luminosity according to Eqs.~\eqref{eq:PLR_shift}--\eqref{eq:deltaMHW}. Because the second rung is calibrated by comparing Cepheid magnitudes in SN~Ia hosts to those in the anchor galaxies, it is the difference in these shifts—not the absolute shift in either rung—that determines the inferred host distances. We do not assume the anchor Cepheids to be screened: their screening state is determined by the proxy field in exactly the same way as for the hosts. Although the geometric distances to the anchors remain unchanged, their Cepheids may still experience a fifth-force-induced brightening. If this brightening is larger (or smaller) than in the SN~Ia hosts, the inferred $H_0$ correspondingly increases (or decreases). A fifth force can therefore move the local distance-ladder determination of $H_0$ either closer to, or further from, the \emph{Planck} value. 

Here, TRGB distances are employed as an optional extension beyond the baseline calibration. When included, they enter the data vector for a fully joint calibration (see Appendix~\ref{sec:AppData} for details). This extension is particularly powerful because TRGB distances shift in the opposite direction to Cepheid distances under a fifth force, providing a complementary constraint.
As shown in Ref.~\cite{PhysRevD.101.129901}, the shift in the TRGB distance to galaxy $i$ is given by
\begin{equation}
\label{eq:Delta_mu_TRGB}
    \Delta \mu_i = \frac{5}{2} \log \left[ 1 - 0.04663 \left( 1 + \frac{\Delta G}{\GN}\right)^{8.389} \right] .
\end{equation}
Accordingly, the TRGB-estimated distance is modified as
\begin{equation}
\label{eq:mu_TRGB}
    \mu_i = \mu_i^\mathrm{TRGB} + \Delta \mu_i - \Delta \mu_\mathrm{anch}
\end{equation}
where $\mu_i^\mathrm{TRGB}$ are the tabulated distance moduli from one of the four TRGB datasets that we use (\Fold, \A, \Lnew, \Fnew). When TRGB data are included, each dataset is used separately in a unified calibration together with the SH0ES Cepheid–SN~Ia distance ladder. The term $\mu_\mathrm{anch}$ denotes the corresponding anchor distance.\\

\noindent \textbf{Third rung.} If the progenitor white dwarf of a SN~Ia is unscreened, the enhanced gravitational coupling modifies the Chandrasekhar mass, thereby shifting the intrinsic luminosity of the resulting supernova \cite{Wright:2017rsu,Desmond_2019,Ruchika:2023ugh}.
However, due to screening effects, compact objects, like white dwarfs, generically experience negligible modifications to gravity. 
For reference, the typical mean density of a white dwarf exceeds that of a typical Cepheid by roughly 11 orders of magnitude. Given this stark contrast, we follow \D and assume throughout that the SNe~Ia are unaffected by a fifth force.

\section{The fifth-force model}
While the transition between screened and unscreened regimes can in principle take various forms,
Ref.~\cite{Hogas:2023pjz} showed that a rapid transition in $\relG$ maximizes the impact on the inferred $H_0$.
Motivated by this, and following \D, we model the transition as step-like in the present work,
\begin{equation}
    \label{eq:relG_model}
    \frac{\Delta G}{\GN} = \left\lbrace \begin{array}{ll}
    0, & p \geq p_0 \\
    \relGmax, & p < p_0
    \end{array} \right.
\end{equation}
where $\relGmax$ denotes the unscreened value of the fifth-force enhancement, and $p$ is the proxy field value (i.e., $\log |\Phi|$, $\log |a|$, or $\log |K|$). The parameter $p_0$ characterizes the transition scale from screened to unscreened.
The two model parameters $(p_0, \relGmax)$, collectively denoted by $\theta$, fully specify the fifth-force model.

\section{Statistical methodology}
\label{sec:StatisticalMethods}
All three rungs of the distance ladder are calibrated jointly within a Bayesian framework. The observed data vector $\mathbf{y}$ is fixed (imported from Ref.~\cite{MarcusSH0ESGitHub}), while the model prediction includes both the standard linear contribution $L \mathbf{q}$ \cite{Riess:2021jrx} and an additive fifth-force correction $\Delta \mathbf{m}(\theta)$ that depends on the nonlinear model parameters $\theta$.
The likelihood is given by
\begin{equation}
\label{eq:like}
    \mathcal{L}(\mathbf{q},\theta) = \frac{\exp{\left[ - \frac{1}{2} \big( \mathbf{y} - \mathbf{y}_\mathrm{th}(\mathbf{q}, \theta) \big)^T C^{-1} \big( \mathbf{y} - \mathbf{y}_\mathrm{th}(\mathbf{q}, \theta) \big) \right]}}{\sqrt{\det C}}
\end{equation}
with
\begin{equation}
    \mathbf{y}_\mathrm{th}(\mathbf{q}, \theta) \equiv L \mathbf{q} + \Delta \mathbf{m}(\theta)
\end{equation}
and where $\mathbf{q}$ denotes the linear model parameters, arranged in a vector, and $\Delta \mathbf{m}(\theta)$ is the modification to the model-prediction induced by the fifth force, that is, the shifts in the Cepheid PL relation and, optionally, the TRGB distances, see Eqs.~\eqref{eq:PLR_shift}--\eqref{eq:mu_TRGB}.
The fifth-force parameters $\theta = (p_0, \relGmax)$ enter the model nonlinearly through $\Delta \mathbf{m}(\theta)$, whereas the parameters in $\mathbf{q}$ appear only as coefficients in $L\mathbf{q}$.
The explicit expressions for $L$, $C$, $\mathbf{y}$, and $\mathbf{q}$ are provided in Appendix~\ref{sec:AppData}.

In summary, our model includes 46 linear parameters, $\mathbf{q}$, corresponding to the standard set used in the SH0ES distance-ladder  \cite{Riess:2021jrx}. These are the distance moduli to the 37 SN~Ia host galaxies ($\mu_i$), the geometric anchor distances to N4258 and the LMC ($\mu_\mathrm{N4258}$, $\mu_\mathrm{LMC}$), the fiducial absolute magitude of the Cepheids ($\MHW$), the distance to M31 ($\mu_\mathrm{M31}$), the slope of the Cepheid PL relation ($b_W$), the Cepheid metallicity–luminosity dependence ($Z_W$), the fiducial absolute magnitude of SNe~Ia ($M_B$), the zero-point offset between ground-based and HST photometry ($\Delta \mathrm{zp}$) and the dimensionless Hubble constant parameter $5 \log H_0$.
Additionally, we have the two nonlinear fifth-force parameters $p_0$ and $\relGmax$.

We use Markov Chain Monte Carlo (MCMC) sampling to infer the posterior distribution of the two nonlinear fifth-force parameters, $\theta = (p_0, \relGmax)$, while analytically marginalizing over the 46 linear parameters $\mathbf{q}$. This yields an effective likelihood $\mathcal{L}(\theta)$ that depends only on the fifth-force parameters; for further detail see Appendix~\ref{sec:AppMarg}.

As discussed in Section~\ref{sec:Influence}, the value of $B_i$ depends on the Cepheid's mass and whether it is on its second or third instability-strip crossing. This crossing number is unknown, so for each unscreened Cepheid we treat it as a binary nuisance parameter with two possible values (second vs. third crossing).
For a given Cepheid, we estimate its mass using \cite{Anderson:2016txx}
\begin{equation}
    \log (M_i / M_\odot) = 0.7705  + 0.2754 \, \p_i .
\end{equation}
This leaves two possible values of $B_i$ depending on the instability crossing, listed in Tab.~\ref{tab:CephB}. Since we do not know which crossing applies, the likelihood sums over both possibilities, effectively marginalizing over the two assignments for each unscreened Cepheid.
The explicit marginalization procedure and its implementation are provided in Appendix~\ref{sec:AppMarg}; see also Ref.~\cite{Hogas:2025lwk} for a comprehensive description.

To sample the posterior distribution of the model parameters, we use the affine-invariant ensemble MCMC sampler implemented in the Python package \texttt{emcee} \cite{foreman2013emcee}. Convergence is monitored using the estimated autocorrelation time, $\tau$: each chain is run to a length of at least $100 \, \tau_\text{max}$, where $\tau_\text{max}$ is the maximum of $\tau$ over the parameters. The initial burn-in phase is removed by discarding the first $2 \, \tau_\text{max}$ samples of each chain. We sample the two nonlinear fifth-force parameters, $p_0$ and $\relGmax$, using 16 parallel walkers whose initial positions are drawn randomly from their respective prior ranges.

We assume uniform priors in $\theta = (p_0, \relGmax)$. Although the full two-dimensional parameter space is infinite, we leverage its asymptotic behavior in defining the prior window. For the screening threshold parameter $p_0$, we set the uniform prior range to $[\min\limits_i \left( p_i - 5 \sigma_{p,i} \right), \; \max\limits_i \left( p_i + 5 \sigma_{p,i} \right)]$ where the minimum and maximum are taken over all galaxies.\footnote{Since $p_0$ corresponds to $\log |\Phi_0|$, $\log |a_0|$, or $\log |K_0|$, the prior is log-uniform in the proxy field itself.} Here, $p_i$ denotes the (log) proxy field value for galaxy $i$ and $\sigma_{p,i}$ is its associated uncertainty. This range ensures that, at the lower end, all galaxies are screened, while at the higher end, all are unscreened. Beyond these limits, all galaxies retain the same screening state, so the distance ladder predictions remain unchanged. Expanding the prior would therefore not probe any new physical effects.
For $\relGmax$, we adopt a uniform prior over $[0, 0.3]$: we find that no posterior would extend to larger $\relGmax$ values if allowed, except when all galaxies are screened, in which case the fifth force is inactive anyway.

\section{Results}
\label{sec:results}
\subsection{Baseline SH0ES calibration with a fifth force}
Using the methodology described above, we infer the posterior distribution of the fifth-force parameters $\theta = (p_0,\relGmax)$ while analytically marginalizing over the standard (linear) SH0ES calibration parameters.
Recall that $p_0$ denotes the threshold proxy-field value which determines the transition between screened and unscreened regimes.
The posterior inference for $\theta$ is carried out separately for each of the three proxy-field models and for five different values of the cutoff radius $\Rmax$.
A complete list of figures and results for all model and dataset configurations is publicly available at \cite{DistLadFifthForceGitHub}.

\begin{figure}
    \centering
    \includegraphics[width=1\linewidth]{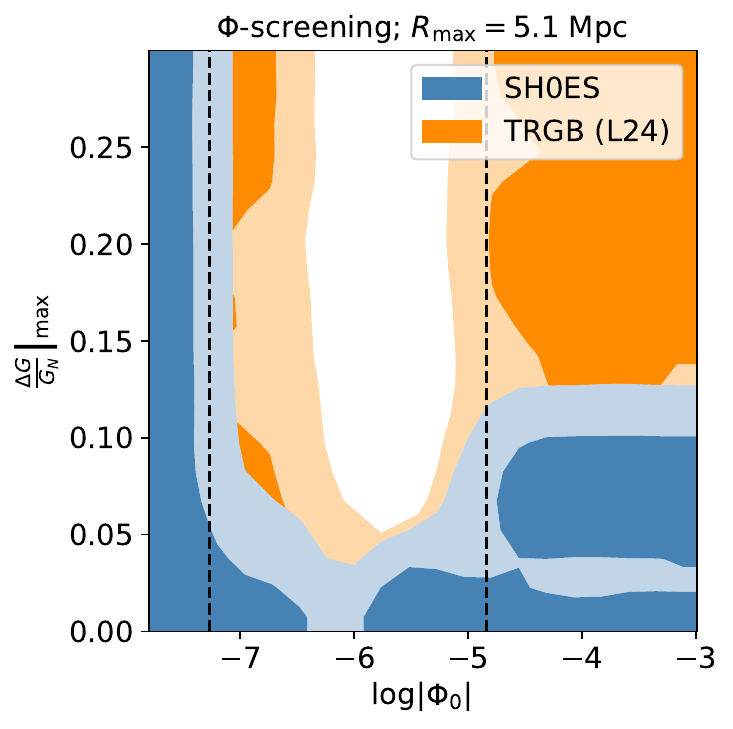}
    \caption{Posterior distributions of the fifth-force parameters for the $\Phi$-screening model with $R_\mathrm{max} = 5.1 \, \mpc$, as an example. Shown are the $68 \, \%$ and $95 \, \%$ confidence contours for the recalibrated SH0ES dataset, and the same when the first two rungs of the distance ladder are replaced by the TRGB distances of L24, yielding weaker constraints. For the SH0ES calibration, the posterior is concentrated near regions consistent with $\Delta G / G_\mathrm{N} = 0$, showing no significant evidence for a fifth-force deviation. The left (right) vertical line indicates the $p_0$-value below (above) which all galaxies are screened (unscreened).}
    \label{fig:PosteriorExample}
\end{figure}

Across all screening models and dataset combinations, we find no statistically significant evidence for a fifth force in the distance ladder. The posterior distributions for the fifth-force parameters are concentrated to regions where $\relG = 0$, that is, no fifth-force effect.
In regions where one or more galaxies are unscreened, the allowed unscreened fifth-force strength $\relG$ is at most a few percent.
See Fig.~\ref{fig:PosteriorExample} for an example.
Thus, there is no evidence for an $H_0$ different from the standard calibration.

As shown in Fig.~\ref{fig:PosteriorExample}, the TRGB-only calibration—where the first two rungs of the SH0ES Cepheid-based distance ladder are replaced by TRGB distances—produces weaker constraints than the Cepheid-based SH0ES calibration.
The same qualitative behavior is found across all screened-gravity models we analyze.
This is not because individual TRGB distances are much less precise than Cepheid-based distances (their uncertainties are comparable), but because the TRGB-only ladder contains fewer anchor and host galaxies and therefore probes a much narrower range of environments in the proxy fields.\footnote{For example, the \Fnew dataset contains only one anchor galaxy and 10 SN host galaxies, compared with the Cepheid-based ladder, which includes three anchor galaxies and 42 SN host galaxies. For completeness, it should be noted that \Fnew contains additional hosts beyond those overlapping with the SH0ES distance ladder, but since we are mainly concerned with a unified Cepheid+TRGB calibration, we restrict the analysis to the common set.}
Consequently, incorporating TRGB distances to the SH0ES distance ladder yields only a marginal increase in constraining power.

The standard SH0ES calibration is naturally recovered in the limits $\relGmax \to 0$ or $p_0 \to -\infty$. In the former case, the fifth force becomes negligible everywhere, while in the latter, all galaxies are screened.
In the opposite limit, $p_0 \to \infty$, all galaxies are unscreened. At first glance, one might expect this to reproduce the standard calibration as well, resulting in a U-shaped confidence contour, since the fifth-force correction would affect all galaxies equally and thus leave the relative distances between anchors and SN~Ia host galaxies unchanged.
However, the fifth-force correction to Cepheid luminosities is not uniform across the Cepheids.
The magnitude shift, Eq.~\eqref{eq:deltaMHW}, depends on the parameter $B_i$, which varies from star to star due to its dependence on the Cepheid’s mass and instability-strip crossing. Thus, even when every galaxy is unscreened, individual Cepheids receive different luminosity shifts, effectively introducing additional scatter around the PL relation, reducing the overall fit quality.
As a result, the likelihood flattens out into a plateau around $\relGmax \simeq 0.1$ as we approach large values of $p_0$.

\begin{figure}
    \centering
    \includegraphics[width=1\linewidth]{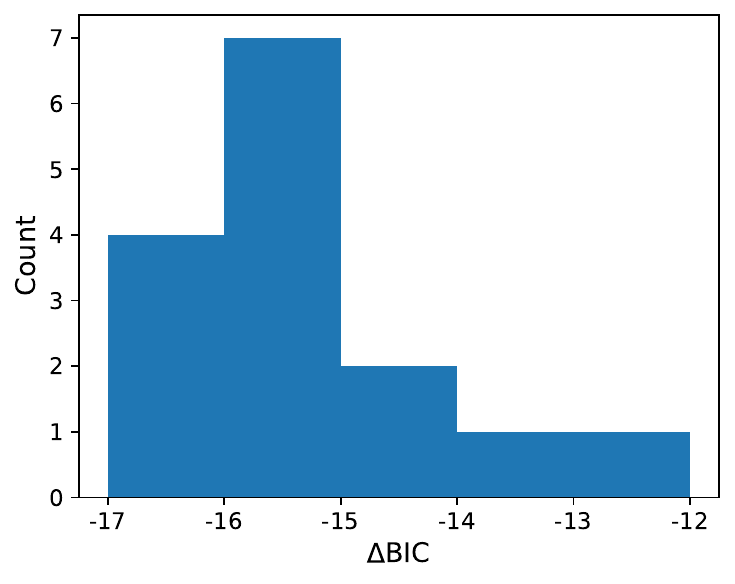}
    \caption{Collection of $\Delta\mathrm{BIC}$ values across all fifth-force models. Each entry corresponds to one combination of screening proxy ($\Phi$, $a$, $K$) and smoothing scale $R_{\max}$. Negative $\Delta\mathrm{BIC}$ values indicate preference for the standard SH0ES calibration without a fifth force. The histogram shows that all fifth-force models lie deep in the region disfavored by the BIC, confirming the absence of statistical support for a fifth-force extension of the distance ladder.}
    \label{fig:DeltaBIC_histogram_noTRGB}
\end{figure}

Model comparison via the Bayesian Information Criterion (BIC) strongly disfavors the fifth-force extensions of the SH0ES distance ladder, with $\Delta \mathrm{BIC} < -12$ in all cases as shown in Fig.~\ref{fig:DeltaBIC_histogram_noTRGB}.
In the Jeffrey–Raftery terminology, this classifies as very strong evidence against a fifth-force model \cite{raftery}.
The Akaike Information Criterion (AIC) tells a consistent but weaker story, as expected from its lower penalty for additional parameters.
A similar conclusion is reached when incorporating TRGB distances in the SH0ES distance ladder.
For the TRGB-only calibration, the preference remains in the same direction but with a reduced statistical significance.

\begin{figure}
    \centering
    \includegraphics[width=1\linewidth]{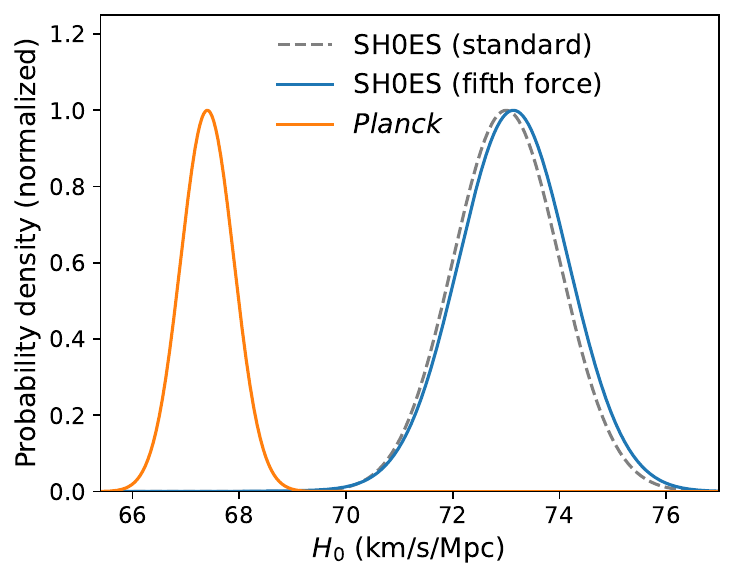}
    \caption{Unified posterior distributions for the Hubble constant $H_0$ obtained by combining results from all 15 screened-gravity models (three proxy fields, five cutoff radii). The result is $H_0 = 73.1 \pm 1.0 \, \hu$. For comparison, the \emph{Planck} estimate is also shown, $H_0 = 67.4 \pm 0.5 \, \hu$, as well as the standard SH0ES calibration without a fifth force, $H_0 = 73.0 \pm 1.0 \, \hu$. The Hubble tension remains at $> 5 \, \sigma$ also in the presence of a fifth force.}
    \label{fig:H0_all_models_noTRGB}
\end{figure}

To assess the impact of a fifth force on $H_0$, we construct a unified posterior over all models.
For each of the 15 screened-gravity models (three proxy fields combined with five cutoff radii), we take the sampled $H_0$ values from the MCMC chain of each model and turn them into a smooth, normalized probability curve defined on a common $H_0$ grid.
Each model is then assigned a weight based on its $\Delta$BIC, using the weight $w_k \propto \exp \left[ \Delta \mathrm{BIC}_k / 2 \right]$ to approximate the posterior model probability of model $k$. This is simply the standard conversion from BIC differences to posterior model probabilities and therefore corresponds to an ordinary probability weighting of the models (a Bayesian model average).
The unified posterior is then obtained by a weighted point-by-point sum of the 15 model-level probability densities. This average represents the overall, model-agnostic fifth-force inference of $H_0$.
The result is shown in Fig.~\ref{fig:H0_all_models_noTRGB}.
From this unified distribution, we infer
\begin{equation}
    H_0 = 73.1 \pm 1.0 \, \hu
\end{equation}
corresponding to a $+0.1 \, \hu$ shift relative to the standard SH0ES calibration without a fifth force. This tiny change leaves the discrepancy with the \emph{Planck} value essentially unchanged, with the tension remaining at \hbox{$> 5 \, \sigma$}.

\subsection{Impact of proxy-field uncertainties}
For each fifth-force model, the proxy-field value assigned to each galaxy is taken to be the median of the externally estimated proxy values listed in Tab.~\ref{tab:ProxyVal}. These values come from the large-scale structure reconstruction of Ref.~\cite{Desmond_2017} and are not inferred from the distance-ladder analysis itself.
To assess the sensitivity of our results to uncertainties in the proxy-field estimates, we repeat the calibration under an intentionally exaggerated ``high-contrast'' scenario. In this test, the proxy values for the anchor galaxies are set to their upper $95 \, \%$ confidence limits, while those for the SN~Ia host galaxies are set to their lower $95 \, \%$ limits—making the anchors maximally screened and the hosts maximally unscreened within the allowed range.
This configuration produces the largest possible downward shift in the inferred $H_0$, and therefore represents the maximal potential reduction of the Hubble tension permitted by the proxy-field uncertainties.

Under this extreme configuration, the results remain qualitatively unchanged. Across all 15 screened-gravity models, model selection continues to disfavor the fifth-force extensions, with $\Delta\mathrm{BIC}$ values ranging from $-16$ to $-10$, that is, strong evidence against the additional parameters. Averaging over all models yields a unified posterior of
\begin{equation}
    H_0 = 73.1 \pm 1.0 \, \hu,
\end{equation}
identical to the baseline result and leaving the Hubble tension essentially unaffected. A small minority of models achieve relatively low best-fit values—two cases fall below $71 \, \hu$: $\Phi$-screening with $\Rmax = 0.4 \, \mpc$ and $a$-screening with $\Rmax = 5.1 \, \mpc$. The lowest occurs for the latter with $H_0 = 70.6 \pm 1.0 \, \hu$. Also here, the modest improvement in fit ($\Delta\chi^2 = -6.7$) is insufficient to offset the additional model complexity, yielding $\Delta\mathrm{BIC} = -10.1$. Thus, even under maximally favorable proxy-field assumptions, there is no sign of screened fifth forces providing a statistically viable path toward lowering $H_0$ to relieve the Hubble tension.

\subsection{Minimum allowed $H_0$}
Even if the distance-ladder data provide no evidence for a fifth force, one may still ask whether such a force could nevertheless \emph{allow} an $H_0$ low enough to be compatible with the early-Universe measurement from \emph{Planck}. To test this, we examine the lowest $H_0$ value permitted at $95 \, \%$ confidence for each model.
Across all models, the minimum allowed $H_0$ shifts by a median of $-0.8 \, \hu$ from the standard (no fifth force) calibration, and even this small change is accompanied by a poorer overall fit.
However, one model stands out: the gravitational-potential proxy ($\Phi$) with cutoff scale $\Rmax = 0.4 \, \mpc$. In this case, the lower $95 \, \%$ bound on $H_0$ reaches $67.8 \, \hu$—a value attained at $(\log |\Phi_0|, \relGmax) = (-8.27, 0.043)$. The model’s best fit on the other hand is $H_0 = 70.6 \, \hu$, achieved at $(\log |\Phi_0|, \relGmax) = (-8.29, 0.021)$. Although significantly lower than the standard SH0ES calibration without a fifth force, model selection strongly disfavors the scenario overall, with $\Delta\mathrm{BIC} = -12.8$. Thus, achieving this low Hubble constant requires adopting a model that is strongly disfavored by the BIC.

\subsection{Including the \emph{Planck} prior}
To quantify the balance between the potential gain of a lower $H_0$ (which would ease the Hubble tension) and the accompanying degradation in the distance-ladder fit, we incorporate the \emph{Planck} measurement of $H_0$ as an external constraint in the likelihood. (This is \emph{not} included in the baseline case, only here.) This allows us to assess whether any screened fifth-force model can achieve a statistically acceptable compromise between improving consistency with \emph{Planck} while still providing a competitive fit to the distance-ladder data.
The answer is that this is not possible: $\Delta\mathrm{BIC}$ remains negative across all proxy and $\Rmax$ combinations, indicating a persistent model preference for standard gravity.

The only exception is the $\Phi$–screening model with $\Rmax = 0.4 \, \mpc$, which attains $\Delta\mathrm{BIC} = +9.5$ once the \emph{Planck} prior on $H_0$ is imposed. In this setting, the model is therefore strongly favored over the standard calibration. However, this support does not persist when the TRGB distances are incorporated into a unified calibration.
Instead, the evidence is substantially weakened and becomes mixed across the TRGB datasets: three cases (\Fold, \A, and \Lnew) yield weak preference against the fifth-force extension, with $\Delta\mathrm{BIC}$ values of $-2.6$, $-0.3$, and $-2.9$, respectively, while one dataset (\Fnew) shows positive support with $\Delta\mathrm{BIC} = +5.0$.

\section{Discussion}
\label{sec:Discussion}
The absence of statistically significant evidence for a fifth force in the distance ladder provides new, astrophysically grounded constraints on screened modified-gravity scenarios.
While solar-system and laboratory experiments already impose stringent limits on deviations from GR in those environments, the present analysis probes a complementary regime: galactic-scale environments with a broad range of external potentials, accelerations, and curvature.
Our analysis shows that screened modified gravity models of the type we consider can produce deviations from standard gravity of at most a few percent across these galaxies—too small to bias the inferred Hubble constant in any significant way.

Our results indicate that fifth forces with screening governed by the environmental gravitational potential, acceleration or curvature are limited in the distance ladder to a few percent of the strength of gravity. This accords with the constraints obtained in \cite{Desmond_2017, Desmond:2020wep} by comparing Cepheid and TRGB distances to common galaxies. It is also at roughly the same level that the strength of gravity in the LMC was probed relative to the Solar System by studying detached eclipsing binaries containing Cepheids in \cite{G_LMC}. Note that specific theories implementing screening have typically been constrained to far greater precision than afforded by these model-agnostic probes; an example is the Hu-Sawicki model of $f(R)$ gravity \cite{Hu_Sawicki}, which the most stringent constraints from \cite{Desmond:2020gzn, Landim} render completely inoperative within galaxies. One would therefore need a new screened theory evading these constraints to have an appreciable effect on the local distance ladder.

An important caveat is that we have analysed only $\Phi$, $a$ and $K$ as screening proxies; this is a subset of the models considered in \cite{Desmond_2019, Desmond:2020wep}, which is itself a subset of all possible screening models. In particular we have considered only external screening here (sourced by the large-scale structure surrounding a test galaxy), while full theories such as chameleons and symmetrons incorporate both environmental and internal screening, sourced by the galaxies' own mass. Other models may therefore have greater success at ameliorating the Hubble tension. An example is the baryon--dark matter interaction model for dark energy of \cite{BDM}, which was found in \cite{Sakstein:2019qgn} to exhibit a novel a screening mechanism dependent on the local dark matter density. This was shown in \cite{Desmond_2019, Desmond:2020wep} to be the most promising model for reducing the distance ladder-inferred $H_0$.

\section{Conclusions}
We have recalibrated the fourth-iteration SH0ES distance ladder in the presence of a screened fifth-force. The modification is parameterized as a two-parameter step function of the environmental proxy $p$, with transition scale $p_0$ and maximum strength $\relGmax$.
By evaluating three alternative proxies ($\log |\Phi|,\log |a|,\log |K|$) across five cutoff scales $\Rmax$, this framework efficiently spans a class of screening mechanisms without committing to any single underlying theory.

Our main conclusions are:

\begin{itemize}
    \item \textbf{No evidence for a fifth force.} Across all models, posteriors concentrate around $\relG=0$.
    Model comparison strongly disfavors fifth-force extensions with $\Delta\mathrm{BIC} < -12$ for all models (Fig.~\ref{fig:DeltaBIC_histogram_noTRGB}).
    
    \item \textbf{The Hubble tension remains.}
    A unified posterior over all screened-gravity models, based on the standard SH0ES Cepheid–SN~Ia calibration, gives
    \begin{equation}
    H_0 = 73.1 \pm 1.0 , \hu,
    \end{equation}
    which is $+0.1 \,\hu$ higher than the SH0ES result obtained without a fifth force (Fig.~\ref{fig:H0_all_models_noTRGB}). This negligible shift leaves the Hubble tension intact at $> 5 \, \sigma$.
    This conclusion holds despite adopting modeling choices designed to maximize the potential impact of a fifth force on the distance ladder—most notably the use of a step-like screening transition and the assumption that the Cepheid core and envelope always share the same screening state.\footnote{This modeling choice does not force Cepheids to be unscreened; rather, it ensures that whenever a galaxy is unscreened, both the core and the envelope contribute maximally to the luminosity shift, thereby maximizing the possible impact of a fifth force on the PL relation.}
    Among the individual models, one case yields a noticeably lower Hubble constant: the $\Phi$-screening model with $\Rmax = 0.4 \, \mpc$, which has a best-fit value of $H_0 = 70.6 \pm 1.0 \, \hu$. However, the improvement in fit is modest ($\Delta \chi^2 = -4.1$) and, once the model complexity is taken into account, it remains strongly disfavored by the BIC, with $\Delta \mathrm{BIC} = -12.8$.
    
    \item \textbf{Proxy-field uncertainties do not hide a large effect.}
    Even under an exaggerated ``high-contrast'' test—assigning anchors and hosts to opposite $95\,\%$ proxy extremes—the resulting $H_0$ remains unchanged from the baseline. Two models yield relatively low best-fit values ($H_0 < 71\,\hu$), the lowest being $a$-screening with $\Rmax = 5.1\,\mathrm{Mpc}$ at $H_0 = 70.6 \pm 1.0\,\hu$, but all are strongly disfavored by model selection ($-16 < \Delta\mathrm{BIC} < -10$). Thus, even extreme proxy assignments do not reveal any viable fifth-force scenario capable of reducing the Hubble tension significantly.

    \item \textbf{Robustness to the inclusion of TRGB data.} As an optional extension to our baseline SH0ES analysis, we incorporated TRGB distances into the same unified calibration—fitting TRGB, Cepheid, and SN~Ia data simultaneously rather than treating TRGB as a separate external check. Even if the TRGB constraints on a fifth force are substantially weaker, due to a smaller sample size, this inclusion is still informative because TRGB and Cepheid distances respond in opposite directions to a fifth force, providing a complementary lever arm. We tested four TRGB datasets—two with HST photometry (\Fold, \A) and two with JWST photometry (\Lnew, \Fnew)—from two independent teams, and in all cases the conclusions remain qualitatively unchanged: the data allow at most percent-level deviations from standard gravity and do not alleviate the Hubble tension. 
    
    \item \textbf{Complementary limits on screened gravity.}
    Our results show that the fractional enhancement of gravity, $\relG$, is constrained to be at most a few percent across all galaxies in the distance ladder. These galaxies span a wide range of gravitational potentials, accelerations, and curvatures, probing environments that are complementary to those tested in the solar system or in laboratory experiments.
\end{itemize}

\begin{acknowledgements}
Thanks to Yukei Murakami for helpful discussions concerning the MCMC code of Ref.~\cite{Riess_2021}, publicly available on GitHub \cite{PantheonPlusData}.

This research utilized the Sunrise HPC facility supported by the Technical Division at the Department of Physics, Stockholm University.

MH and EM acknowledges support from the Swedish Research Council under Dnr VR 2020-03384. HD is supported by a Royal Society University Research Fellowship (grant no. 211046).

The authors used OpenAI's ChatGPT to assist in drafting portions of the manuscript.
\end{acknowledgements}

\section*{Data availability}
All data required to reproduce the results of this work are available at \cite{DistLadFifthForceGitHub}.
The repository includes both original analysis products and externally sourced data sets compiled for convenience. The latter are publicly available from their original authors and are cited in the text.

\appendix
\section{Standard linear model}
\label{sec:AppData}
Our analysis builds on the fourth-iteration SH0ES calibration \cite{Riess:2021jrx}. For convenience, we use a reformatted version of the same dataset, arranged to enable straightforward implementation of alternative calibration schemes \cite{MarcusSH0ESGitHub}.

In the baseline (no fifth-force) case, the calibration is identical in structure to the standard SH0ES analysis and remains linear in the model parameters $\mathbf{q}$. We also allow for two optional extensions:
(i) the inclusion of TRGB distances (one dataset at a time), entering directly into the data vector and calibrated jointly with the Cepheid and SN~Ia data; and
(ii) the addition of an external constraint on $H_0$ from \emph{Planck}. When included, these extensions appear as additional blocks at the end of the vectors in Eqs.~\eqref{eq:y_R22}–\eqref{eq:L_q_R22}.

For the standard calibration without a fifth force, the likelihood takes the form
\begin{equation}
\label{eq:LikeLinear}
    \mathcal{L}(\mathbf{q}) = \frac{\exp{\left[ - \frac{1}{2} \big( \mathbf{y} - L \mathbf{q}\big)^T C^{-1} \big( \mathbf{y} - L \mathbf{q} \big) \right]}}{\sqrt{\det C}}.
\end{equation}
In the main analysis of this work, we then introduce a fifth force. This modifies the model through an additive term $\Delta \mathbf{m}(\theta)$ to the model prediction $L \mathbf{q}$. This term depends nonlinearly on the fifth-force parameters $\theta$ and is incorporated directly into the likelihood as described in Section~\ref{sec:Influence}.

The explicit forms of the data vector $\mathbf{y}$, covariance matrix $C$, design matrix $L$, and linear parameter vector $\mathbf{q}$ used in this baseline calibration are provided below:
\begin{widetext}
    \begin{equation}
    \label{eq:y_R22}
    \mathbf{y} = 
    \begin{array}{ll}
    \left(
    \begin{array}[c]{c}
    
    m^W_{H,\mathrm{M101}} \\

    : \\

    m^W_{H,\mathrm{U9391}} \\
    
    \hline
    
    m^W_{H,\textrm{N4258}} \\
    
    m^W_{H,\textrm{M31}} \\
    
    m^W_{H,\textrm{LMC,GRND}} \\

    m^W_{H,\textrm{LMC,HST}} \\
    
    m^W_{H,\textrm{SMC}} \\
    
    \hline
    
    m_{B,\mathrm{Cal \; SN}} \\

    \hline
    
    m_{B,\mathrm{HF \; SN}} - 5 \log \left[ c z \{ ... \} \right] -25 \\
    
    \hline
    
    M_{H,\textrm{HST}}^W \\
    
    M_{H,{\textrm{Gaia}}}^W \\
    
    Z_{W} \\
    
    0 \\
    
    \mu_\mathrm{N4258}^\mathrm{anch} \\
    
    \mu_\mathrm{LMC}^\mathrm{anch} \\

    \hline

    \mu_i \\

    \hline

    5 \log H_0^{Planck}
    
    \end{array} \right)
    
    &
    
    \begin{array}[c]{@{}l@{\,}l}
    
    \left.
    \begin{array}{c} \vphantom{m^W_{H,\mathrm{hosts}}} \\
    \vphantom{:} \\
    \vphantom{m^W_{H,\mathrm{hosts}}}
    \end{array}
    \right\} & \text{2150 Cepheids in SNIa hosts} \\
    
    \left.
    \begin{array}{c} \vphantom{m^W_{H,\textrm{nh},j}} \\ 
    \vphantom{m^W_{H,\textrm{nh},j}} \\ 
    \vphantom{m^W_{H,\textrm{nh},j}} \\
    \vphantom{m^W_{H,\textrm{nh},j}} \\
    \vphantom{\textrm{\LARGE HELLO}}
    \end{array}
    \right\} & \text{(443 + 55 + 270 + 69 + 143) Cepheids in anchors or non-SNIa hosts\hspace{1.5in}} \\
    
    \left.
    \begin{array}{c}
    \vphantom{m_B^0}
    \end{array}
    \right\} & \text{77 Cal SNe magnitudes} \\

    \left.
    \begin{array}{c}
    \vphantom{.}
    \end{array}
    \right\} & \text{277 HF SNe} \\
    
    \left.
    \begin{array}{c}
    \vphantom{.} \\
    \vphantom{.} \\
    \vphantom{.} \\
    \vphantom{.} \\ 
    \vphantom{.} \\
    \vphantom{.}
    \end{array}
    \right\} & \text{6 External constraints} \\

    \left.
    \begin{array}{c}
    \vphantom{.}
    \end{array}
    \right\} & \text{Optional: distance moduli from TRGB} \\

    \left.
    \begin{array}{c}
    \vphantom{.}
    \end{array}
    \right\} & \text{Optional: $H_0$ from \textit{Planck}}
    \end{array}
    \end{array}
    \end{equation}

    \begin{equation}
    \label{eq:C_R22}
        C = \left(
        \begin{array}{cccccccccccccccccc}
        
        \sy{\sigma_{\rm M101}^2}\!\!\!\! & \td & \sy{Z_{\textrm{cov}}} & \sy{Z_{\textrm{cov}}} & \sy{0} & \sy{0} & \sy{0} & \sy{0} & \sy{0} & \sy{0}  & \sy{0} & \sy{0} & \sy{0} & \sy{0} & \sy{0} & \sy{0} & \sy{0} & \sy{0} \\
        
        : & \rotatebox{45}{:} & : & : & : & : & : & : & : & : & : & : & : & : & : & : & : & : \\
        
        \sy{Z_{\textrm{cov}}} & \td & \sy{\sigma_{\rm U9391}^2}\!\!\!\! & \sy{Z_{\textrm{cov}}} & \sy{0} & \sy{0} & \sy{0} & \sy{0} &\sy{0} & \sy{0} & \sy{0} & \sy{0} & \sy{0} & \sy{0} & \sy{0} & \sy{0} & \sy{0} & \sy{0} \\
        
        \hline
        
        \sy{Z_{\textrm{cov}}} & \td & \sy{Z_{\textrm{cov}}} & \sy{\sigma_{{\rm N4258}}^2}\!\!\!\! & \sy{0} & \sy{0} & \sy{0} & \sy{0}  &\sy{0} & \sy{0} & \sy{0} & \sy{0} & \sy{0} & \sy{0}  & \sy{0} & \sy{0} & \sy{0} & \sy{0} \\
        
        \sy{0} & \td & \sy{0} & \sy{0} & \sy{\sigma_{{\rm M31}}^2}\!\!\!\! & \sy{0} & \sy{0} & \sy{0}  &\sy{0} & \sy{0} & \sy{0} & \sy{0} & \sy{0} & \sy{0}  & \sy{0} & \sy{0} & \sy{0} & \sy{0} \\

        \sy{0} & \td & \sy{0} & \sy{0} & \sy{0} &  \sy{\sigma_{{\rm LMC,GRND}}^2}\!\!\!\! & \sy{10^{-4}} & \sy{0} & \sy{0} & \sy{0} & \sy{0} & \sy{0} & \sy{0} & \sy{0}  & \sy{0} & \sy{0} & \sy{0} & \sy{0} \\

        \sy{0} & \td & \sy{0} & \sy{0} & \sy{0} & \sy{10^{-4}} &  \sy{\sigma_{{\rm LMC,HST}}^2}\!\!\!\! & \sy{0} & \sy{0} & \sy{0} & \sy{0} & \sy{0} & \sy{0} & \sy{0} & \sy{0}  & \sy{0} & \sy{0} & \sy{0} \\

        \sy{0} & \td & \sy{0} & \sy{0} & \sy{0} & \sy{0} & \sy{0} &  \sy{\sigma_{{\rm SMC}}^2}\!\!\!\! & \sy{0} & \sy{0} & \sy{0} & \sy{0} & \sy{0} & \sy{0} & \sy{0}  & \sy{0} & \sy{0} & \sy{0} \\

        \hline
                
        \sy{0} & \td & \sy{0} & \sy{0} & \sy{0} & \sy{0} & \sy{0} & \sy{0} & \sy{\sigma^2_{\textrm{Cal SN}}}\!\!\!\! & \sy{{\rm SN}_{\textrm{cov}}} & \sy{0} & \sy{0} & \sy{0} & \sy{0} & \sy{0} & \sy{0} & \sy{0} & \sy{0} \\
        
        \hline

        \sy{0} & \td & \sy{0} & \sy{0} & \sy{0} & \sy{0} & \sy{0} & \sy{0} & \sy{{\rm SN}_{\textrm{cov}}} & \sy{\sigma^2_{\textsc{HF SN}}}\!\!\!\! & \sy{0} & \sy{0} & \sy{0} & \sy{0} & \sy{0} & \sy{0} & \sy{0} & \sy{0} \\
        
        \hline
        
        \sy{0} & \td & \sy{0} & \sy{0} & \sy{0} & \sy{0} & \sy{0} & \sy{0} & \sy{0} & \sy{0} & \sy{\sigma_{\rm HST}^2} & \sy{0}\!\!\!\! & \sy{0} & \sy{0} & \sy{0}  & \sy{0} & \sy{0} & \sy{0} \\
        
        \sy{0} & \td & \sy{0} & \sy{0} & \sy{0} & \sy{0} & \sy{0} & \sy{0} & \sy{0} &  \sy{0} & \sy{0} & \sy{\sigma_{\rm Gaia}^2} & \sy{0}\!\!\!\! & \sy{0} & \sy{0} & \sy{0} & \sy{0} & \sy{0} \\

        \sy{0} & \td & \sy{0} & \sy{0} & \sy{0} & \sy{0} & \sy{0} & \sy{0} & \sy{0} &  \sy{0} & \sy{0} & \sy{0} & \sy{\sigma_{Z_W}^2} & \sy{0}\!\!\!\! & \sy{0} & \sy{0} & \sy{0} & \sy{0} \\
        
        \sy{0} & \td & \sy{0} & \sy{0} & \sy{0} & \sy{0} & \sy{0} & \sy{0} & \sy{0} & \sy{0} & \sy{0} & \sy{0} & \sy{0} & \sy{\sigma_{\rm grnd}^2}\!\!\!\! & \sy{0} & \sy{0} & \sy{0} & \sy{0} \\
        
        \sy{0} & \td & \sy{0} & \sy{0} & \sy{0} & \sy{0} & \sy{0} & \sy{0} & \sy{0} & \sy{0} & \sy{0} & \sy{0} & \sy{0} & \sy{0} & \sy{\sigma_{\mu,{\rm N4258}}^2}\!\!\!\! & \sy{0} & \sy{0} & \sy{0} \\
        
        \sy{0} & \td & \sy{0} & \sy{0} & \sy{0} & \sy{0} & \sy{0} & \sy{0} &  \sy{0} & \sy{0} & \sy{0} & \sy{0} & \sy{0} & \sy{0} & \sy{0} & \sy{\sigma_{\mu,{\rm LMC}}^2} & \sy{0} & \sy{0} \\

        \hline
        
        \sy{0} & \td & \sy{0} & \sy{0} & \sy{0} & \sy{0} & \sy{0} & \sy{0} &  \sy{0} & \sy{0} & \sy{0} & \sy{0} & \sy{0} & \sy{0} & \sy{0} & \sy{0} & \sy{\sigma_{\mu_i}^2} & \sy{0} \\

        \hline

        \sy{0} & \td & \sy{0} & \sy{0} & \sy{0} & \sy{0} & \sy{0} & \sy{0} &  \sy{0} & \sy{0} & \sy{0} & \sy{0} & \sy{0} & \sy{0} & \sy{0} & \sy{0} & \sy{0} & \sy{\sigma_{5 \log H_0}^2}
        \end{array}
        \right)
    \end{equation}

    \begin{equation}
    \label{eq:L_q_R22}
        L =
        \left(
        \begin{array}[c]{ccccccclclcc}
        
        1 & \td & 0 & 0 & 1 & 0 & 0 & [P]_\mathrm{M101} & 0 & \oh_\mathrm{M101} & 0 & 0 \\
        
        : & \rotatebox{45}{:} & : & : & : & : & :   & : & : & : & : & : \\  
        
        0 & \td & 1 & 0 & 1 & 0 & 0 & [P]_\mathrm{U9391} & 0 & \oh_\mathrm{U9391} & 0 & 0 \\
        
        \hline
        
        0 & \td & 0 & 1 & 1 & 0 & 0 & [P]_\mathrm{N4258} & 0 & \oh_{\textrm{N4258}} & 0 & 0 \\
        
        0 &\td & 0 & 0 & 1 & 0 & 1 & [P]_\mathrm{M31} & 0 & \oh_{\textrm{M31}}   & 0 & 0 \\
        
        0 & \td & 0 & 0 & 1 & 1 & 0 & [P]_\mathrm{LMC,GRND} & 0 & \oh_{\textrm{LMC,GRND}} & 1 & 0 \\

        0 & \td & 0 & 0 & 1 & 1 & 0 & [P]_\mathrm{LMC,HST} & 0 & \oh_{\textrm{LMC,HST}} & 0 & 0 \\

        0 & \td & 0 & 0 & 1 & 1 & 0 & [P]_\mathrm{SMC} & 0 & \oh_{\textrm{SMC}}    & 1 & 0 \\
        
        \hline
        
        1 & \td & 0 & 0 & 0 & 0 & 0 & 0 & 1 & 0 & 0 & 0 \\
        
        : & \rotatebox{45}{:} & : & : & : & : & : & : & : & : & : & : \\  
        
        0 & \td & 1 & 0 & 0 & 0 & 0 & 0 & 1 & 0 & 0 & 0 \\
        
        \hline

        0 & \td & 0 & 0 & 0 & 0 & 0 & 0 & 1 & 0 & 0 & -1 \\

        : & \rotatebox{45}{:} & : & : & : & : & : & : & : & : & : & : \\
        
        0 & \td & 0 & 0 & 0 & 0 & 0 & 0 & 1 & 0 & 0 & -1\\

        \hline
        
        0 & \td & 0 & 0 & 1 & 0 & 0 & 0 & 0 & 0 & 0 & 0 \\
        
        0 & \td & 0 & 0 & 1 & 0 & 0 & 0 & 0 & 0 & 0 & 0 \\

        0 & \td & 0 & 0 & 0 & 0 & 0 & 0 & 0 & 1 & 0 & 0 \\
        
        0 & \td & 0 & 0 & 0 & 0 & 0 & 0 & 0 & 0 & 1 & 0 \\
        
        0 & \td & 0 & 1 & 0 & 0 & 0 & 0 & 0 & 0 & 0 & 0 \\
        
        0 & \td & 0 & 0 & 0 & 1 & 0 & 0 & 0 & 0 & 0 & 0 \\

        \hline
        
        1 & \td & 0 & 0 & 0 & 0 & 0 & 0 & 0 & 0 & 0 & 0 \\
        
        : & \rotatebox{45}{:} & : & : & : & : & : & : & : & : & : & : \\  
        
        0 & \td & 1 & 0 & 0 & 0 & 0 & 0 & 0 & 0 & 0 & 0 \\
        
        \hline

        0 & \td & 0 & 0 & 0 & 0 & 0 & 0 & 0 & 0 & 0 & 1
        
        \end{array}
        \right), \quad \quad
        \mathbf{q} = 
        \left(
        \begin{array}{c}
            \mu_\mathrm{M101} \\
            : \\
            \mu_\mathrm{U9391} \\
            \hline
            \mu_\mathrm{N4258} \\
            \MHW \\
            \mu_\mathrm{LMC} \\
            \mu_\mathrm{M31} \\
            b_W \\
            M_B \\
            Z_W \\
            \Delta \mathrm{zp} \\
            5 \log H_0
        \end{array}
        \right) .
    \end{equation}
\end{widetext}
Substituting these expressions into the likelihood in Eq.~\eqref{eq:LikeLinear} and performing a standard least-squares fit yields $H_0 = 73.0 \pm 1.0 \, \hu$, identical to the published SH0ES result \cite{Riess:2021jrx}, as expected.

The expression within curly brackets in Eq.~\eqref{eq:y_R22} appears on the left-hand side of Eq.~\eqref{eq:mBHFSN}. In the covariance matrix \eqref{eq:C_R22}, the terms $\sigma_\mathrm{M101}, \dots, \sigma_\mathrm{U9391}$, $\sigma_\mathrm{N4258}$, $\sigma_\mathrm{M31}$, $\sigma_\mathrm{LMC}$, $\sigma_\mathrm{SMC}$, $\sigma_\mathrm{Cal \; SN}$, and $\sigma_\mathrm{HF \; SN}$ denote the individual covariance blocks associated with each galaxy or supernova sample, including their off-diagonal elements. Additionally, $Z_\mathrm{cov}$ accounts for correlations among Cepheid host galaxies that may arise from a common systematic uncertainty in metallicity.

The distance modulus to N4258 is determined from water megamaser observations at the galaxy’s center \cite{Reid_2019}, while that to the LMC is obtained via detached eclipsing binaries \cite{Pietrzy_ski_2019}. These geometric distance estimates are incorporated as external constraints in Eqs.~\eqref{eq:y_R22}--\eqref{eq:L_q_R22}.

For the MW Cepheids, geometric distances are derived from parallax measurements. The calibration utilizes two independent parallax datasets: one from HST \cite{Riess:2014uga,Riess_2018} and another from \emph{Gaia} EDR3 \citep{Riess_2021}. The MW Cepheid photometry and periods are provided in \cite{Riess_2018,Riess:2018byc,Riess_2021}. In Eqs.~\eqref{eq:y_R22}--\eqref{eq:L_q_R22}, MW Cepheid data enter as external constraints on $\MHW$, with separate constraints for HST and \emph{Gaia} EDR3, and as an external constraint on $Z_W$, derived from the \emph{Gaia} EDR3 Cepheids.

To ensure photometric consistency and avoid zero-point systematics in the computation of the Wesenheit magnitudes, the SH0ES compilation \cite{Riess:2021jrx} employs Cepheid photometry measured uniformly in the HST WFC3 system. Magnitude, color, and period measurements are obtained from Refs.~\cite{Kato:2007zze,Soszynski:2008kd,Macri:2014xpa,Kodric:2018hpc,Riess_2019,2019wfc..rept....1R,2021ApJ...920...84L,Riess:2021jrx} and the Cepheid metallicities from Refs.~\cite{Romaniello:2008yh,2018A&A...619A...8G,Romaniello:2021vht,Riess:2021jrx}.
A small subset of Cepheids are observed from the ground rather than with HST. This includes the “GRND” subset of LMC Cepheids, as well as the SMC Cepheids, whose photometry is sourced from Refs.~\cite{Kato:2007zze,Soszynski:2008kd,Macri:2014xpa}. To incorporate these in a uniform HST WFC3-based analysis, their ground-based photometry is transformed following the prescription of Ref.~\cite{Riess:2016jrr}. This introduces an additional zero-point parameter, $\Delta \mathrm{zp}$, which captures the residual offset between the ground-based and HST systems and is treated as a free parameter subject to the constraint $\Delta \mathrm{zp} = 0 \pm 0.10$.

For SNe~Ia, the redshifts, magnitudes, and covariance matrix are adopted from the Pantheon+ dataset \cite{Scolnic:2021amr}.

\section{Marginalization}
\label{sec:AppMarg}
The likelihood was given in Eq.~\eqref{eq:like}.
We write the theoretical model-prediction as the standard linear contribution $\sum_j L_{ij} q_j$, plus a fifth-force correction $\Delta m_i(\theta, s)$,
\begin{equation}
    y^\mathrm{th}_i(\theta,q,s) = \sum_j L_{ij} \, q_j + \Delta m_i(\theta,s),
\end{equation}
where $L_{ij}$ is the design matrix, $q_i$ is the vector of linear model parameters, and $\theta$ denotes the nonlinear fifth-force parameters ($p_0$ and $\relGmax$).
The fifth-force correction can be expressed as
\begin{equation}
    \Delta m_i(\theta, s) = \Delta_i(\theta) + s_i d_i(\theta)
\end{equation}
where $\Delta_i(\theta)$ is a deterministic mean-shift, and $s_i d_i(\theta)$ is the stochastic residual from the unknown Cepheid instability strip crossing.
When the index $i$ corresponds to a Cepheid data point, $\Delta_i$ is a mean-shift in the Wesenheit magnitude
\begin{equation}
\label{eq:Delta}
    \Delta_i(\theta) = -2.5 \left( \frac{A}{2} + \overline{B}_i \right) \log \left( 1 + \frac{\Delta G}{\GN} \right)
\end{equation}
with $\overline{B}_i \equiv (B_{i,2} + B_{i,3} )/2$ being the mean $B_i$ between the second and third instability strip crossing, cf. Eq.~\eqref{eq:deltaMHW}.

Note that $B_i$ is a function of both the Cepheid's mass and whether it is on the second or third crossing of the instability strip, as shown in Tab.~\ref{tab:CephB}. The mass can be estimated using \cite{Anderson:2016txx}
\begin{equation}
    \log (M_i / M_\odot) = 0.7705  + 0.2754 \, \p_i .
\end{equation}
The stochastic residual $s_i d_i$, with $s_i = \pm 1$, is the residual shift in the Wesenheit magnitude depending on whether the Cepheid belongs to the second ($s_i = +1$) or third ($s_i = -1$) crossing of the instability strip.
Here,
\begin{equation}
    d_i(\theta) = 2.5 \, \Delta B_i \log \left( 1 + \frac{\Delta G}{\GN} \right)
\end{equation}
with $\Delta B_i \equiv (B_{i,2} - B_{i,3}) / 2$. The $\theta$-dependence enters via $\relG$. 

In vector notation, the likelihood can now be written
\begin{widetext}
    \begin{equation}
    \label{eq:Like_theta_q_s}
        \mathcal{L}(\mathbf{q}, \mathbf{s}, \theta) = \frac{ \exp\left[ - \tfrac12 \big( \mathbf{y} - L \mathbf{q} - \mathbf{\Delta}(\theta) - \mathbf{s} \odot \mathbf{D}(\theta) \big)^T C^{-1} \big( \mathbf{y} - L \mathbf{q} - \mathbf{\Delta}(\theta) - \mathbf{s} \odot \mathbf{D}(\theta) \big) \right] }{ \sqrt{\det C} } .
    \end{equation}
\end{widetext}
Here,
\begin{subequations}
    \begin{align}
        \mathbf{s} &\equiv \left( \begin{array}{c}
            s_1  \\
            : \\
            s_N
        \end{array} \right), \\
        \mathbf{s} \odot \mathbf{D} &\equiv \left( \begin{array}{c}
             s_1 d_1 \\
             : \\
             s_N d_N
        \end{array} \right).
    \end{align}
\end{subequations}

\subsection{Linear parameters}
To marginalize over the linear parameters, we introduce the posterior covariance of the linear parameters, $\Sigma$, and the Schur complement projector, $P$,
\begin{subequations}
    \begin{align}
        \Sigma &\equiv \big(L^T C^{-1} L\big)^{-1},\\
        P &\equiv C^{-1} - C^{-1} L \Sigma L^T C^{-1}.
    \end{align}
\end{subequations}
Assuming uniform priors on the linear parameters $\mathbf{q}$, we integrate out the linear parameters and get the marginalized likelihood\footnote{Strictly speaking, since galaxies populate three-dimensional space approximately uniformly rather than uniformly in distance modulus, a uniform prior in the latter is not fully accurate \cite{Desmond:2025ggt}. In the present analysis, however, we follow the standard SH0ES calibration and assume uniform priors on the distance moduli.}
\begin{widetext}
    \begin{equation}
        \mathcal{L}(\mathbf{s}, \theta) = \sqrt{\frac{\det \Sigma}{\det C}} \exp \left[ -\frac12 \left( \mathbf{y} - \mathbf{\Delta} - \mathbf{s} \odot \mathbf{D} \right)^T P \left( \mathbf{y} - \mathbf{\Delta} - \mathbf{s} \odot \mathbf{D} \right) \right]
\end{equation}
\end{widetext}
Note that the normalization factor $\sqrt{ \det \Sigma / \det C }$ is important if the intrinsic PL scatter is fitted as a model parameter.

\subsection{Instability crossing number}
Marginalizing over the unknown instability-strip crossings, $\mathbf{s}$, the Cepheid likelihood becomes a sum over all possible assignments of $s_i=\pm 1$:  
\begin{equation}
\label{eq:Like_theta}
    \mathcal{L}(\theta) =
    \sum_{\mathbf{s} \in \{\pm 1\}^N} \mathcal{L}(\mathbf{s}, \theta),
\end{equation}
where $N$ is the number of Cepheids.
In Eq.~\eqref{eq:Like_theta}, we have assumed equal probability for a Cepheid to be on either the second or third instability-strip crossing, so that each term in the sum carries equal weight. 

We can factorize out the $s$-independent part of Eq.~\eqref{eq:Like_theta} so that
\begin{equation}
\label{eq:Like_theta_2}
    \mathcal{L}(\theta) = \sqrt{\frac{\det \Sigma}{\det C}} e^{-\tfrac12 \overline{\mathbf{y}}^T P \overline{\mathbf{y}} } \sum_{\mathbf{s} \in \{\pm 1\}^N} e^{ \tfrac12 \mathbf{s}^T J \mathbf{s} + \mathbf{s}^T \mathbf{h} }.
\end{equation}
Here, we have defined $\overline{\mathbf{y}}$ as the mean-shifted data vector, $\overline{\mathbf{y}} \equiv \mathbf{y} - \mathbf{\Delta}$, which thus depends on $\theta$ (but not $\mathbf{s}$). 
The interaction matrix $J$ and field $\mathbf{h}$ are defined component-wise as
\begin{equation}
\label{eq:Def_J_h}
    J_{ij} \equiv - d_i P_{ij} d_j, 
    \qquad 
    h_i \equiv - d_i \sum_j P_{ij} \overline{y}_j .
\end{equation}
Importantly, $J$ and $\mathbf{h}$ depend on the fifth-force parameters $\theta$ but not on $\mathbf{s}$. 

The sum in Eq.~\eqref{eq:Like_theta_2} is mathematically identical to the partition function of the Ising model with spins $s_i=\pm 1$, couplings $J_{ij}$, and external magnetic field $h_i$.
A positive off-diagonal element of $J_{ij}$ corresponds to a ferromagnetic coupling between the spin $i$ and $j$ particles whereas a negative off-diagonal corresponds to an antiferromagnetic coupling.
With the distance-ladder covariance matrix (tabulated in Ref.~\cite{MarcusSH0ESGitHub}), ferromagnetic couplings dominate.
The connection between binary model parameters and the Ising model is explored in detail in Ref.~\cite{Hogas:2025lwk}.

In practice, evaluating the likelihood \eqref{eq:Like_theta} exactly is impossible: the sum contains $2^N$ terms, and the projected covariance $P$ introduces off-diagonal couplings in $J$.
However, approximating $P$ as diagonal, the sum factorizes and can be carried out analytically, most conveniently expressed
\begin{equation}
\label{eq:DeltaLCeph}
   \ln \sum_{\mathbf{s} \in \{\pm 1\}^N} e^{ \tfrac12 \mathbf{s}^T J \mathbf{s} + \mathbf{s}^T \mathbf{h} } \simeq \frac12 \mathrm{Tr}\,J 
   + \sum_i \ln \!\big[2\cosh(h_i)\big].
\end{equation}
The neglected off-diagonal terms represent correlations between Cepheids (or, in the Ising-model language, pairwise interactions). These terms enter only at second order in the perturbation to the likelihood—that is, as a correction to the correction—so their numerical impact is expected to be small.
A quantitative assessment of this statement is provided in Ref.~\cite{Hogas:2025lwk}, where the next-order improvement (a mean-field treatment retaining off-diagonal couplings) was compared directly to the paramagnetic approximation in controlled test cases. In those tests, the impact of the mean-field correction was tiny, indicating that the off-diagonal terms contribute only subdominant corrections in realistic regimes.

The final log-likelihood, marginalized over the linear and binary parameters, reads
\begin{widetext}
    \begin{equation}
    \label{eq:loglike_final}
        \ln \mathcal{L}(\theta) = - \frac12 \left[ \overline{\mathbf{y}}^T(\theta) P \, \overline{\mathbf{y}}(\theta) + \ln \det C - \ln \det \Sigma - \mathrm{Tr} \, J(\theta) - 2 \sum_i \ln \left( 2 \cosh h_i(\theta) \right) \right] .
    \end{equation}
\end{widetext}
We emphasize that the approximation $P \simeq \mathrm{diag}(P)$ was used only for approximating the sum in Eq.~\eqref{eq:DeltaLCeph}. In the final expression for the marginalized log-likelihood, Eq.~\eqref{eq:loglike_final}, $P$ is the full projected, non-diagonal covariance matrix.

\clearpage
\bibliography{bibliography}{}

@ARTICLE{NPP,
       author = {{Baker}, Tessa and {Barreira}, Alexandre and {Desmond}, Harry and {Ferreira}, Pedro and {Jain}, Bhuvnesh and {Koyama}, Kazuya and {Li}, Baojiu and {Lombriser}, Lucas and {Nicola}, Andrina and {Sakstein}, Jeremy and {Schmidt}, Fabian},
        title = "{The Novel Probes Project -- Tests of Gravity on Astrophysical Scales}",
      journal = {arXiv e-prints},
         year = 2019,
        month = aug,
          eid = {arXiv:1908.03430},
        pages = {arXiv:1908.03430},
          doi = {10.48550/arXiv.1908.03430},
archivePrefix = {arXiv},
       eprint = {1908.03430},
 primaryClass = {astro-ph.CO},
       adsurl = {https://ui.adsabs.harvard.edu/abs/2019arXiv190803430B}
}

@article{Brax:2021wcv,
    author = "Brax, Philippe and Casas, Santiago and Desmond, Harry and Elder, Benjamin",
    title = "{Testing Screened Modified Gravity}",
    eprint = "2201.10817",
    archivePrefix = "arXiv",
    primaryClass = "gr-qc",
    doi = "10.3390/universe8010011",
    journal = "Universe",
    volume = "8",
    number = "1",
    pages = "11",
    year = "2021"
}

@ARTICLE{Hu_Sawicki,
       author = {{Hu}, Wayne and {Sawicki}, Ignacy},
        title = "{Models of f(R) cosmic acceleration that evade solar system tests}",
      journal = {\prd},
         year = 2007,
        month = sep,
       volume = {76},
       number = {6},
          eid = {064004},
        pages = {064004},
          doi = {10.1103/PhysRevD.76.064004},
archivePrefix = {arXiv},
       eprint = {0705.1158},
 primaryClass = {astro-ph},
       adsurl = {https://ui.adsabs.harvard.edu/abs/2007PhRvD..76f4004H}
}

@ARTICLE{Landim,
       author = {{Landim}, Ricardo G. and {Desmond}, Harry and {Koyama}, Kazuya and {Penny}, Samantha},
        title = "{Testing screened modified gravity with SDSS-IV-MaNGA}",
      journal = {\mnras},
         year = 2024,
        month = oct,
       volume = {534},
       number = {1},
        pages = {349-360},
          doi = {10.1093/mnras/stae2096},
archivePrefix = {arXiv},
       eprint = {2407.08825},
 primaryClass = {astro-ph.CO},
       adsurl = {https://ui.adsabs.harvard.edu/abs/2024MNRAS.534..349L}
}

@ARTICLE{G_LMC,
       author = {{Desmond}, Harry and {Sakstein}, Jeremy and {Jain}, Bhuvnesh},
        title = "{Five percent measurement of the gravitational constant in the Large Magellanic Cloud}",
      journal = {\prd},
         year = 2021,
        month = jan,
       volume = {103},
       number = {2},
          eid = {024028},
        pages = {024028},
          doi = {10.1103/PhysRevD.103.024028},
archivePrefix = {arXiv},
       eprint = {2012.05028},
 primaryClass = {astro-ph.CO},
       adsurl = {https://ui.adsabs.harvard.edu/abs/2021PhRvD.103b4028D}
}

@ARTICLE{BDM,
       author = {{Berezhiani}, Lasha and {Khoury}, Justin and {Wang}, Junpu},
        title = "{Universe without dark energy: Cosmic acceleration from dark matter-baryon interactions}",
      journal = {\prd},
         year = 2017,
        month = jun,
       volume = {95},
       number = {12},
          eid = {123530},
        pages = {123530},
          doi = {10.1103/PhysRevD.95.123530},
archivePrefix = {arXiv},
       eprint = {1612.00453},
 primaryClass = {hep-th},
       adsurl = {https://ui.adsabs.harvard.edu/abs/2017PhRvD..95l3530B}
}

@ARTICLE{Clifton,
       author = {{Clifton}, Timothy and {Ferreira}, Pedro G. and {Padilla}, Antonio and {Skordis}, Constantinos},
        title = "{Modified gravity and cosmology}",
      journal = {\physrep},
         year = 2012,
        month = mar,
       volume = {513},
       number = {1},
        pages = {1-189},
          doi = {10.1016/j.physrep.2012.01.001},
archivePrefix = {arXiv},
       eprint = {1106.2476},
 primaryClass = {astro-ph.CO},
       adsurl = {https://ui.adsabs.harvard.edu/abs/2012PhR...513....1C}
}

@article{Desmond_2019,
   title={Local resolution of the Hubble tension: The impact of screened fifth forces on the cosmic distance ladder},
   volume={100},
   ISSN={2470-0029},
   url={http://dx.doi.org/10.1103/PhysRevD.100.043537},
   DOI={10.1103/physrevd.100.043537},
   number={4},
   journal={Physical Review D},
   publisher={American Physical Society (APS)},
   author={Desmond, Harry and Jain, Bhuvnesh and Sakstein, Jeremy},
   year={2019},
   month={Aug},
   note = {\hypertarget{Desmond_2019}{}}
}

@article{PhysRevD.101.129901,
  title = {Erratum: Local resolution of the Hubble tension: The impact of screened fifth forces on the cosmic distance ladder [Phys. Rev. D 100, 043537 (2019)]},
  author = {Desmond, Harry and Jain, Bhuvnesh and Sakstein, Jeremy},
  journal = {Phys. Rev. D},
  volume = {101},
  issue = {12},
  pages = {129901},
  numpages = {3},
  year = {2020},
  month = {Jun},
  publisher = {American Physical Society},
  doi = {10.1103/PhysRevD.101.129901},
  url = {https://link.aps.org/doi/10.1103/PhysRevD.101.129901}
}

@article{Desmond_2017,
   title={Reconstructing the gravitational field of the local Universe},
   volume={474},
   ISSN={1365-2966},
   url={http://dx.doi.org/10.1093/mnras/stx3062},
   DOI={10.1093/mnras/stx3062},
   number={3},
   journal={Monthly Notices of the Royal Astronomical Society},
   publisher={Oxford University Press (OUP)},
   author={Desmond, Harry and Ferreira, Pedro G and Lavaux, Guilhem and Jasche, Jens},
   year={2017},
   month={Nov},
   pages={3152–3161}
}

@article{Riess:2016jrr,
      author         = "Riess, Adam G. and others",
      title          = "{A 2.4\% Determination of the Local Value of the Hubble
                        Constant}",
      journal        = "\apj",
      volume         = "826",
      year           = "2016",
      number         = "1",
      pages          = "56",
      doi            = "10.3847/0004-637X/826/1/56",
      eprint         = "1604.01424",
      archivePrefix  = "arXiv",
      primaryClass   = "astro-ph.CO",
      SLACcitation   = "%%CITATION = ARXIV:1604.01424;%%"
}

@article{Riess_2021,
  author    = {Riess, Adam G. and Casertano, Stefano and Yuan, Wenlong and Bowers, J. Bradley and Macri, Lucas and Zinn, Joel C. and Scolnic, Dan},
  title     = {Cosmic Distances Calibrated to 1\% Precision with Gaia EDR3 Parallaxes and Hubble Space Telescope Photometry of 75 Milky Way Cepheids Confirm Tension with $\Lambda$CDM},
  journal   = {\apj},
  year      = {2021},
  volume    = {908},
  pages     = {L6},
  month     = {Feb},
  issn      = {2041-8213},
  doi       = {10.3847/2041-8213/abdbaf},
  publisher = {American Astronomical Society},
  url       = {http://dx.doi.org/10.3847/2041-8213/abdbaf},
}

@article{Planck2020,
  author    = {Aghanim, N. and Akrami, Y. and Ashdown, M. and Aumont, J. and Baccigalupi, C. and Ballardini, M. and Banday, A. J. and Barreiro, R. B. and Bartolo, N. and et al.},
  title     = {Planck 2018 results},
  journal   = {\aap},
  year      = {2020},
  volume    = {641},
  pages     = {A6},
  month     = {Sep},
  issn      = {1432-0746},
  doi       = {10.1051/0004-6361/201833910},
  publisher = {EDP Sciences},
  url       = {http://dx.doi.org/10.1051/0004-6361/201833910},
}

@article{Riess_2019,
  author    = {Riess, Adam G. and Casertano, Stefano and Yuan, Wenlong and Macri, Lucas M. and Scolnic, Dan},
  title     = {Large Magellanic Cloud Cepheid Standards Provide a 1\% Foundation for the Determination of the Hubble Constant and Stronger Evidence for Physics beyond $\Lambda$CDM},
  journal   = {\apj},
  year      = {2019},
  volume    = {876},
  number    = {1},
  pages     = {85},
  month     = {May},
  issn      = {1538-4357},
  doi       = {10.3847/1538-4357/ab1422},
  publisher = {American Astronomical Society},
  url       = {http://dx.doi.org/10.3847/1538-4357/ab1422},
}

@article{Reid_2019,
  author    = {Reid, M. J. and Pesce, D. W. and Riess, A. G.},
  title     = {An Improved Distance to NGC 4258 and Its Implications for the Hubble Constant},
  journal   = {\apj},
  year      = {2019},
  volume    = {886},
  number    = {2},
  pages     = {L27},
  month     = {Nov},
  issn      = {2041-8213},
  doi       = {10.3847/2041-8213/ab552d},
  publisher = {American Astronomical Society},
  url       = {http://dx.doi.org/10.3847/2041-8213/ab552d},
}

@article{Freedman_2019,
  author    = {Freedman, Wendy L. and Madore, Barry F. and Hatt, Dylan and Hoyt, Taylor J. and Jang, In Sung and Beaton, Rachael L. and Burns, Christopher R. and Lee, Myung Gyoon and Monson, Andrew J. and Neeley, Jillian R. and et al.},
  title     = {The Carnegie-Chicago Hubble Program. VIII. An Independent Determination of the Hubble Constant Based on the Tip of the Red Giant Branch},
  journal   = {\apj},
  year      = {2019},
  volume    = {882},
  number    = {1},
  pages     = {34},
  month     = {Aug},
  issn      = {1538-4357},
  doi       = {10.3847/1538-4357/ab2f73},
  publisher = {American Astronomical Society},
  url       = {http://dx.doi.org/10.3847/1538-4357/ab2f73},
  note = {\hypertarget{Freedman_2019}{}}
}

@Misc{2019wfc..rept....1R,
  author       = {{Riess}, A.~G. and {Narayan}, Gautham and {Calamida}, Annalisa},
  title        = {{Calibration of the WFC3-IR Count-rate Nonlinearity, Sub-percent Accuracy for a Factor of a Million in Flux}},
  howpublished = {Space Telescope WFC Instrument Science Report},
  month        = jan,
  year         = {2019},
  adsurl       = {https://ui.adsabs.harvard.edu/abs/2019wfc..rept....1R},
  pages        = {1},
}

@article{Pietrzy_ski_2019,
   title={A distance to the Large Magellanic Cloud that is precise to one per cent},
   volume={567},
   ISSN={1476-4687},
   url={http://dx.doi.org/10.1038/s41586-019-0999-4},
   DOI={10.1038/s41586-019-0999-4},
   number={7747},
   journal={Nature},
   publisher={Springer Science and Business Media LLC},
   author={Pietrzyński, G. and Graczyk, D. and Gallenne, A. and Gieren, W. and Thompson, I. B. and Pilecki, B. and Karczmarek, P. and Górski, M. and Suchomska, K. and Taormina, M. and et al.},
   year={2019},
   month={Mar},
   pages={200–203}
}

@article{Riess:2021jrx,
    author = "Riess, Adam G. and others",
    title = "{A Comprehensive Measurement of the Local Value of the Hubble Constant with 1 km s$^{−1}$ Mpc$^{−1}$ Uncertainty from the Hubble Space Telescope and the SH0ES Team}",
    eprint = "2112.04510",
    archivePrefix = "arXiv",
    primaryClass = "astro-ph.CO",
    doi = "10.3847/2041-8213/ac5c5b",
    journal = "Astrophys. J. Lett.",
    volume = "934",
    number = "1",
    pages = "L7",
    year = "2022"
}

@article{Will:2014kxa,
    author = "Will, Clifford M.",
    title = "{The Confrontation between General Relativity and Experiment}",
    eprint = "1403.7377",
    archivePrefix = "arXiv",
    primaryClass = "gr-qc",
    doi = "10.12942/lrr-2014-4",
    journal = "Living Rev. Rel.",
    volume = "17",
    pages = "4",
    year = "2014"
}

@article{Sakstein:2019qgn,
    author = "Sakstein, Jeremy and Desmond, Harry and Jain, Bhuvnesh",
    title = "{Screened Fifth Forces Mediated by Dark Matter--Baryon Interactions: Theory and Astrophysical Probes}",
    eprint = "1907.03775",
    archivePrefix = "arXiv",
    primaryClass = "astro-ph.CO",
    doi = "10.1103/PhysRevD.100.104035",
    journal = "Phys. Rev. D",
    volume = "100",
    number = "10",
    pages = "104035",
    year = "2019"
}

@article{Wright:2017rsu,
    author = "Wright, Bill S. and Li, Baojiu",
    title = "{Type Ia supernovae, standardizable candles, and gravity}",
    eprint = "1710.07018",
    archivePrefix = "arXiv",
    primaryClass = "astro-ph.CO",
    doi = "10.1103/PhysRevD.97.083505",
    journal = "Phys. Rev. D",
    volume = "97",
    number = "8",
    pages = "083505",
    year = "2018"
}

@article{Hogas:2023vim,
    author = {H\"og\r{a}s, Marcus and M\"ortsell, Edvard},
    title = "{Impact of symmetron screening on the Hubble tension: New constraints using cosmic distance ladder data}",
    eprint = "2303.12827",
    archivePrefix = "arXiv",
    primaryClass = "astro-ph.CO",
    doi = "10.1103/PhysRevD.108.024007",
    journal = "Phys. Rev. D",
    volume = "108",
    number = "2",
    pages = "024007",
    year = "2023"
}

@article{Desmond:2020wep,
    author = "Desmond, Harry and Sakstein, Jeremy",
    title = "{Screened fifth forces lower the TRGB-calibrated Hubble constant too}",
    eprint = "2003.12876",
    archivePrefix = "arXiv",
    primaryClass = "astro-ph.CO",
    doi = "10.1103/PhysRevD.102.023007",
    journal = "Phys. Rev. D",
    volume = "102",
    number = "2",
    pages = "023007",
    year = "2020"
}

@article{Khoury:2013tda,
    author = "Khoury, Justin",
    title = "{Les Houches Lectures on Physics Beyond the Standard Model of Cosmology}",
    eprint = "1312.2006",
    archivePrefix = "arXiv",
    primaryClass = "astro-ph.CO",
    month = "12",
    year = "2013"
}

@article{Joyce:2014kja,
    author = "Joyce, Austin and Jain, Bhuvnesh and Khoury, Justin and Trodden, Mark",
    title = "{Beyond the Cosmological Standard Model}",
    eprint = "1407.0059",
    archivePrefix = "arXiv",
    primaryClass = "astro-ph.CO",
    doi = "10.1016/j.physrep.2014.12.002",
    journal = "Phys. Rept.",
    volume = "568",
    pages = "1--98",
    year = "2015"
}

@article{Ruchika:2023ugh,
    author = "Ruchika and Rathore, Himansh and Roy Choudhury, Shouvik and Rentala, Vikram",
    title = "{A gravitational constant transition within cepheids as supernovae calibrators can solve the Hubble tension}",
    eprint = "2306.05450",
    archivePrefix = "arXiv",
    primaryClass = "astro-ph.CO",
    doi = "10.1088/1475-7516/2024/06/056",
    journal = "JCAP",
    volume = "06",
    pages = "056",
    year = "2024"
}

@article{Desmond:2020gzn,
    author = "Desmond, Harry and Ferreira, Pedro G.",
    title = "{Galaxy morphology rules out astrophysically relevant Hu-Sawicki $f(R)$ gravity}",
    eprint = "2009.08743",
    archivePrefix = "arXiv",
    primaryClass = "astro-ph.CO",
    doi = "10.1103/PhysRevD.102.104060",
    journal = "Phys. Rev. D",
    volume = "102",
    number = "10",
    pages = "104060",
    year = "2020"
}

@article{Anand:2021sum,
    author = "Anand, Gagandeep S. and Tully, R. Brent and Rizzi, Luca and Riess, Adam G. and Yuan, Wenlong",
    title = "{Comparing Tip of the Red Giant Branch Distance Scales: An Independent Reduction of the Carnegie-Chicago Hubble Program and the Value of the Hubble Constant}",
    eprint = "2108.00007",
    archivePrefix = "arXiv",
    primaryClass = "astro-ph.CO",
    doi = "10.3847/1538-4357/ac68df",
    journal = "Astrophys. J.",
    volume = "932",
    number = "1",
    pages = "15",
    year = "2022",
    note = {\hypertarget{Anand:2021sum}{}}
}

@article{foreman2013emcee,
       author = {{Foreman-Mackey}, Daniel and {Hogg}, David W. and {Lang}, Dustin and {Goodman}, Jonathan},
        title = "{emcee: The MCMC Hammer}",
      journal = {Publ. Astron. Soc. Pac.},
         year = 2013,
       volume = {125},
       number = {925},
        pages = {306},
          doi = {10.1086/670067},
archivePrefix = {arXiv},
       eprint = {1202.3665},
 primaryClass = {astro-ph.IM},
       adsurl = {https://ui.adsabs.harvard.edu/abs/2013PASP..125..306F}
}

@article{Hogas:2023pjz,
    author = {H\"og\r{a}s, Marcus and M\"ortsell, Edvard},
    title = "{Hubble tension and fifth forces}",
    eprint = "2309.01744",
    archivePrefix = "arXiv",
    primaryClass = "astro-ph.CO",
    doi = "10.1103/PhysRevD.108.124050",
    journal = "Phys. Rev. D",
    volume = "108",
    number = "12",
    pages = "124050",
    year = "2023"
}

@article{Hogas:2024qlt,
    author = {H{\"o}g{\r{a}}s, Marcus and M{\"o}rtsell, Edvard},
    title = "{Reassessing the Cepheid-based distance ladder: implications for the Hubble constant}",
    eprint = "2412.07840",
    archivePrefix = "arXiv",
    primaryClass = "astro-ph.CO",
    doi = "10.1093/mnras/staf308",
    journal = "Mon. Not. Roy. Astron. Soc.",
    volume = "538",
    number = "2",
    pages = "883--906",
    year = "2025"
}

@article{Scolnic:2021amr,
    author = "Scolnic, Dan and others",
    title = "{The Pantheon+ Analysis: The Full Data Set and Light-curve Release}",
    eprint = "2112.03863",
    archivePrefix = "arXiv",
    primaryClass = "astro-ph.CO",
    doi = "10.3847/1538-4357/ac8b7a",
    journal = "ApJ",
    fjournal = "Astrophys. J.",
    volume = "938",
    number = "2",
    pages = "113",
    year = "2022"
}

@article{Leavitt:1912zz,
    author = "Leavitt, Henrietta S. and Pickering, Edward C.",
    title = "{Periods of 25 Variable Stars in the Small Magellanic Cloud}",
    reportNumber = "HCO-CIRC-173",
    journal = "Harv. Obs. Circ.",
    volume = "173",
    pages = "1--3",
    year = "1912"
}

@ARTICLE{Leavitt:1907,
       author = {{Leavitt}, Henrietta S.},
        title = "{1777 variables in the Magellanic Clouds}",
        journal = {Ann. Harv. Coll. Obs.},
        fjournal = {Annals of Harvard College Observatory},
         year = 1907,
        month = jan,
       volume = {60},
        pages = {87-108.3},
       adsurl = {https://ui.adsabs.harvard.edu/abs/1907AnHar..60...87L}
}

@article{Madore:1991yf,
    author = "Madore, Barry F. and Freedman, Wendy L.",
    title = "{The Cepheid distance scale}",
    doi = "10.1086/132911",
    journal = "PASP",
    fjournal = "Publ. Astron. Soc. Pac.",
    volume = "103",
    pages = "933--957",
    year = "1991"
}

@article{Riess:2014uga,
    author = "Riess, Adam G. and Casertano, Stefano and Anderson, Jay and Mackenty, John and Filippenko, Alexei V.",
    title = "{Parallax Beyond a Kiloparsec from Spatially Scanning the Wide Field Camera 3 on the Hubble Space Telescope}",
    eprint = "1401.0484",
    archivePrefix = "arXiv",
    primaryClass = "astro-ph.IM",
    doi = "10.1088/0004-637X/785/2/161",
    journal = "ApJ",
    fjournal = "Astrophys. J.",
    volume = "785",
    pages = "161",
    year = "2014"
}

@article{Riess_2018,
doi = {10.3847/1538-4357/aaadb7},
url = {https://dx.doi.org/10.3847/1538-4357/aaadb7},
year = {2018},
month = {mar},
publisher = {The American Astronomical Society},
volume = {855},
number = {2},
pages = {136},
author = {Adam G. Riess and Stefano Casertano and Wenlong Yuan and Lucas Macri and Jay Anderson and John W. MacKenty and J. Bradley Bowers and Kelsey I. Clubb and Alexei V. Filippenko and David O. Jones and Brad E. Tucker},
title = {New Parallaxes of Galactic Cepheids from Spatially Scanning the Hubble Space Telescope: Implications for the Hubble Constant},
journal = {ApJ},
fjournal = {The Astrophysical Journal}
}

@article{Riess:2018byc,
    author = "Riess, Adam G. and others",
    title = "{Milky Way Cepheid Standards for Measuring Cosmic Distances and Application to Gaia DR2: Implications for the Hubble Constant}",
    eprint = "1804.10655",
    archivePrefix = "arXiv",
    primaryClass = "astro-ph.CO",
    doi = "10.3847/1538-4357/aac82e",
    journal = "ApJ",
    fjournal = "Astrophys. J.",
    volume = "861",
    number = "2",
    pages = "126",
    year = "2018"
}

@article{Romaniello:2008yh,
    author = "Romaniello, M. and Primas, F. and Mottini, M. and Pedicelli, S. and Lemasle, B. and Bono, G. and Francois, P. and Groenewegen, M. A. T. and Laney, C. D.",
    title = "{The influence of chemical composition on the properties of Cepheid stars. II-The iron content}",
    eprint = "0807.1196",
    archivePrefix = "arXiv",
    primaryClass = "astro-ph",
    doi = "10.1051/0004-6361:20065661",
    journal = "Astron. Astrophys.",
    fjournal = "Astron. Astrophys.",
    volume = "488",
    pages = "731",
    year = "2008"
}

@ARTICLE{2018A&A...619A...8G,
       author = {{Groenewegen}, M.~A.~T.},
        title = "{The Cepheid period-luminosity-metallicity relation based on Gaia DR2 data}",
        journal = {Astron. Astrophys.},
        fjournal = {Astron.Astrophys.},
         year = 2018,
        month = nov,
       volume = {619},
          eid = {A8},
        pages = {A8},
          doi = {10.1051/0004-6361/201833478},
archivePrefix = {arXiv},
       eprint = {1808.05796},
 primaryClass = {astro-ph.SR},
       adsurl = {https://ui.adsabs.harvard.edu/abs/2018A&A...619A...8G}
}

@article{Romaniello:2021vht,
    author = "Romaniello, Martino and others",
    title = "{The iron and oxygen content of LMC Classical Cepheids and its implications for the extragalactic distance scale and Hubble constant - Equivalent width analysis with Kurucz stellar atmosphere models}",
    eprint = "2110.08860",
    archivePrefix = "arXiv",
    primaryClass = "astro-ph.CO",
    doi = "10.1051/0004-6361/202142441",
    journal = "Astron. Astrophys.",
    fjournal = "Astron. Astrophys.",
    volume = "658",
    pages = "A29",
    year = "2022",
    note = "[Erratum: Astron.Astrophys. 662, C1 (2022)]"
}

@article{Kato:2007zze,
    author = "Kato, Daisuke and others",
    title = "{The IRSF Magellanic Clouds Point Source Catalog}",
    doi = "10.1093/pasj/59.3.615",
    journal = "PASJ",
    fjournal = "Publ. Astron. Soc. Jap.",
    volume = "59",
    pages = "615--641",
    year = "2007"
}

@article{Soszynski:2008kd,
    author = "Soszynski, I. and Poleski, R. and Udalski, A. and Szymanski, M. K. and Kubiak, M. and Pietrzynski, G. and Wyrzykowski, L. and Szewczyk, O. and Ulaczyk, K.",
    title = "{The Optical Gravitational Lensing Experiment. The OGLE-III Catalog of Variable Stars. I. Classical Cepheids in the Large Magellanic Cloud}",
    eprint = "0808.2210",
    archivePrefix = "arXiv",
    primaryClass = "astro-ph",
    journal = "Acta Astron.",
    volume = "58",
    pages = "163",
    year = "2008"
}

@article{Macri:2014xpa,
    author = "Macri, Lucas M. and Ngeow, Chow-Choong and Kanbur, Shashi M. and Mahzooni, Salma and Smitka, Michael T.",
    title = "{Large Magellanic Cloud Near-Infrared Synoptic Survey. I. Cepheid variables and the calibration of the Leavitt Law}",
    eprint = "1412.1511",
    archivePrefix = "arXiv",
    primaryClass = "astro-ph.SR",
    doi = "10.1088/0004-6256/149/4/117",
    journal = "AJ",
    fjournal = "Astron. J.",
    volume = "149",
    pages = "117",
    year = "2015"
}

@article{Kodric:2018hpc,
    author = "Kodric, Mihael and Riffeser, Arno and Seitz, Stella and Hopp, Ulrich and Snigula, Jan and Goessl, Claus and Koppenhoefer, Johannes and Bender, Ralf",
    title = "{M31 PAndromeda Cepheid sample observed in four HST bands}",
    eprint = "1807.08753",
    archivePrefix = "arXiv",
    primaryClass = "astro-ph.GA",
    doi = "10.3847/1538-4357/aad4a1",
    journal = "ApJ",
    fjournal = "Astrophys. J.",
    volume = "864",
    number = "1",
    pages = "59",
    year = "2018"
}

@ARTICLE{2021ApJ...920...84L,
       author = {{Li}, Siyang and {Riess}, Adam G. and {Busch}, Michael P. and {Casertano}, Stefano and {Macri}, Lucas M. and {Yuan}, Wenlong},
        title = "{A Sub-2\% Distance to M31 from Photometrically Homogeneous Near-infrared Cepheid Period-Luminosity Relations Measured with the Hubble Space Telescope}",
        journal = {ApJ},
        fjournal = {\apj},
         year = 2021,
        month = oct,
       volume = {920},
       number = {2},
          eid = {84},
        pages = {84},
          doi = {10.3847/1538-4357/ac1597},
archivePrefix = {arXiv},
       eprint = {2107.08029},
 primaryClass = {astro-ph.CO},
       adsurl = {https://ui.adsabs.harvard.edu/abs/2021ApJ...920...84L}
}

@article{Freedman:2024eph,
    author = "Freedman, Wendy L. and Madore, Barry F. and Hoyt, Taylor J. and Jang, In Sung and Lee, Abigail J. and Owens, Kayla A.",
    title = "{Status Report on the Chicago-Carnegie Hubble Program (CCHP): Measurement of the Hubble Constant Using the Hubble and James Webb Space Telescopes}",
    eprint = "2408.06153",
    archivePrefix = "arXiv",
    primaryClass = "astro-ph.CO",
    doi = "10.3847/1538-4357/adce78",
    journal = "Astrophys. J.",
    volume = "985",
    number = "2",
    pages = "203",
    year = "2025",
    note = {\hypertarget{Freedman:2024eph}{}}
}

@article{Li:2024pjo,
    author = "Li, Siyang and Anand, Gagandeep S. and Riess, Adam G. and Casertano, Stefano and Yuan, Wenlong and Breuval, Louise and Macri, Lucas M. and Scolnic, Daniel and Beaton, Rachael and Anderson, Richard I.",
    title = "{Tip of the Red Giant Branch Distances with JWST. II. I-band Measurements in a Sample of Hosts of 10 Type Ia Supernova Match HST Cepheids}",
    eprint = "2408.00065",
    archivePrefix = "arXiv",
    primaryClass = "astro-ph.CO",
    doi = "10.3847/1538-4357/ad84f3",
    journal = "Astrophys. J.",
    volume = "976",
    number = "2",
    pages = "177",
    year = "2024",
    note = {\hypertarget{Li:2024pjo}{}}
}

@article{Anderson:2016txx,
    author = {Anderson, Richard I. and Saio, Hideyuki and Ekstr{\"o}m, Sylvia and Georgy, Cyril and Meynet, Georges},
    title = "{On the Effect of Rotation on Populations of Classical Cepheids II. Pulsation Analysis for Metallicities 0.014, 0.006, and 0.002}",
    eprint = "1604.05691",
    archivePrefix = "arXiv",
    primaryClass = "astro-ph.SR",
    doi = "10.1051/0004-6361/201528031",
    journal = "Astron. Astrophys.",
    volume = "591",
    pages = "A8",
    year = "2016"
}

@article{Hogas:2025lwk,
    author = {H{\"o}g{\r{a}}s, Marcus and M{\"o}rtsell, Edvard},
    title = "{Analytic Marginalization over Binary Variables in Physics Data}",
    eprint = "2510.21912",
    archivePrefix = "arXiv",
    primaryClass = "astro-ph.CO",
    month = "10",
    year = "2025"
}

@misc{PantheonPlusData,
  author       = "{Pantheon+ Collaboration}",
  title        = "{Pantheon+ data release}",
  howpublished = "\url{https://github.com/PantheonPlusSH0ES/DataRelease}",
  note         = "{{GitHub} repository, accessed 2025-10-31}",
  year         = {2022}
}

@misc{MarcusSH0ESGitHub,
  author       = "{Marcus Högås}",
  title        = "{Cepheid Distance Ladder Data}",
  howpublished = "\url{https://github.com/marcushogas/Cepheid-Distance-Ladder-Data}",
  note         = "{{GitHub} repository}",
  year         = {2024}
}

@article{raftery,
 ISSN = {00811750, 14679531},
 URL = {http://www.jstor.org/stable/271063},
 author = {Adrian E. Raftery},
 journal = {Sociological Methodology},
 pages = {111--163},
 publisher = {[American Sociological Association, Wiley, Sage Publications, Inc.]},
 title = {Bayesian Model Selection in Social Research},
 urldate = {2024-03-01},
 volume = {25},
 year = {1995}
}

@misc{DistLadFifthForceGitHub,
  author       = "{Marcus Högås}",
  title        = "{Distance Ladder Fifth Force}",
  howpublished = "\url{https://github.com/marcushogas/Distance-Ladder-Fifth-Force}",
  note         = "{{GitHub} repository}",
  year         = {2025}
}

@article{Jasche:2018oym,
    author = "Jasche, Jens and Lavaux, Guilhem",
    title = "{Physical Bayesian modelling of the non-linear matter distribution: new insights into the Nearby Universe}",
    eprint = "1806.11117",
    archivePrefix = "arXiv",
    primaryClass = "astro-ph.CO",
    doi = "10.1051/0004-6361/201833710",
    journal = "Astron. Astrophys.",
    volume = "625",
    pages = "A64",
    year = "2019"
}

@article{Lavaux:2015tsa,
    author = "Lavaux, Guilhem and Jasche, Jens",
    title = "{Unmasking the Masked Universe: the 2M++ catalogue through Bayesian eyes}",
    eprint = "1509.05040",
    archivePrefix = "arXiv",
    primaryClass = "astro-ph.CO",
    doi = "10.1093/mnras/stv2499",
    journal = "Mon. Not. Roy. Astron. Soc.",
    volume = "455",
    number = "3",
    pages = "3169--3179",
    year = "2016"
}

@article{Desmond:2025ggt,
    author = "Desmond, Harry and Stiskalek, Richard and Najera, Jose Antonio and Banik, Indranil",
    title = "{The subtle statistics of the distance ladder: On the distance prior and selection effects}",
    eprint = "2511.03394",
    archivePrefix = "arXiv",
    primaryClass = "astro-ph.CO",
    month = "11",
    year = "2025"
}

@article{H0DN:2025lyy,
    author = "Casertano, Stefano and others",
    collaboration = "H0DN",
    title = "{The Local Distance Network: a community consensus report on the measurement of the Hubble constant at 1{\%} precision}",
    eprint = "2510.23823",
    archivePrefix = "arXiv",
    primaryClass = "astro-ph.CO",
    month = "10",
    year = "2025"
}

@article{Riess:2009pu,
    author = "Riess, Adam G. and others",
    title = "{A Redetermination of the Hubble Constant with the Hubble Space Telescope from a Differential Distance Ladder}",
    eprint = "0905.0695",
    archivePrefix = "arXiv",
    primaryClass = "astro-ph.CO",
    doi = "10.1088/0004-637X/699/1/539",
    journal = "Astrophys. J.",
    volume = "699",
    pages = "539--563",
    year = "2009"
}

@article{Riess:2011yx,
    author = "Riess, Adam G. and Macri, Lucas and Casertano, Stefano and Lampeitl, Hubert and Ferguson, Henry C. and Filippenko, Alexei V. and Jha, Saurabh W. and Li, Weidong and Chornock, Ryan",
    title = "{A 3{\%} Solution: Determination of the Hubble Constant with the Hubble Space Telescope and Wide Field Camera 3}",
    eprint = "1103.2976",
    archivePrefix = "arXiv",
    primaryClass = "astro-ph.CO",
    doi = "10.1088/0004-637X/732/2/129",
    journal = "Astrophys. J.",
    volume = "730",
    pages = "119",
    year = "2011",
    note = "[Erratum: Astrophys.J. 732, 129 (2011)]"
}
\bibliographystyle{apsrev4-1}

\end{document}